\documentclass[twocolumn,english,prb,showpacs]{revtex4-1}
\usepackage{epsfig}
\usepackage{graphicx}
\usepackage{amsfonts}
\usepackage[figuresright]{rotating}
\usepackage{amssymb}
\usepackage{amsmath}
\usepackage{amsmath}
\usepackage{graphicx}
\usepackage{hyperref}
\usepackage{color}
\usepackage{soul}
\usepackage{tabularx}
\usepackage{array}
\usepackage{floatrow}
\usepackage{subfig}

\newcolumntype{P}[1]{>{\centering\arraybackslash}p{#1}}

\usepackage{amsmath,amssymb,amsfonts}	
\usepackage{color}
\usepackage{amsmath}
\usepackage{amsfonts}
\usepackage{amssymb}
\usepackage{graphicx}
\usepackage{slashed}            
\usepackage{bbm}                
\usepackage{xspace}				

\usepackage{tikz}
\usetikzlibrary{arrows,shapes}
\usetikzlibrary{trees}
\usetikzlibrary{matrix,arrows} 				
\usetikzlibrary{positioning}				
\usetikzlibrary{calc,through}				
\usetikzlibrary{decorations.pathreplacing}  
\usepackage{pgffor}							

\usetikzlibrary{decorations.pathmorphing}	
\usetikzlibrary{decorations.markings}
\tikzset{
    vector/.style={decorate, decoration={snake}, draw},
	provector/.style={decorate, decoration={snake,amplitude=2.5pt}, draw},
	antivector/.style={decorate, decoration={snake,amplitude=-2.5pt}, draw},
    fermion/.style={draw=black, postaction={decorate},
        decoration={markings,mark=at position .55 with {\arrow[draw=black]{>}}}},
    fermionbar/.style={draw=black, postaction={decorate},
        decoration={markings,mark=at position .55 with {\arrow[draw=black]{<}}}},
    fermionnoarrow/.style={draw=black},
    gluon/.style={decorate, draw=black,
        decoration={coil,amplitude=4pt, segment length=5pt}},
    scalar/.style={dashed,draw=black, postaction={decorate},
        decoration={markings,mark=at position .55 with {\arrow[draw=black]{>}}}},
    scalarbar/.style={dashed,draw=black, postaction={decorate},
        decoration={markings,mark=at position .55 with {\arrow[draw=black]{<}}}},
    scalarnoarrow/.style={dashed,draw=black},
    electron/.style={draw=black, postaction={decorate},
        decoration={markings,mark=at position .55 with {\arrow[draw=black]{>}}}},
	bigvector/.style={decorate, decoration={snake,amplitude=4pt}, draw},
}

\tikzstyle{block} = [draw, rectangle, 
    minimum height=3em, minimum width=6em]

\usepackage{hyperref}			

\DeclareCaptionJustification{justified}{\leftskip=0pt \rightskip=0pt \parfillskip=0pt plus 1fil}

\begin{document}

\floatsetup[figure]{style=plain,subcapbesideposition=top}

\title{Floquet band structure of a semi-Dirac system}
\author{Qi Chen}
\email[]{chenqi0805@gmail.com}
\author{Liang Du}
\author{Gregory A. Fiete}
\affiliation{Department of Physics, The University of Texas at
Austin, Austin, TX, 78712, USA}
\begin{abstract}

In this work we use Floquet-Bloch theory to study the influence of circularly and linearly polarized light on two-dimensional band structures with semi-Dirac band touching points, taking the anisotropic nearest neighbor hopping model on the honeycomb lattice as an example. We find circularly polarized light opens a gap and induces a band inversion to create a finite Chern number in the two-band model. By contrast, linearly polarized light can either open up a gap (polarized in the quadratically dispersing direction) or split the semi-Dirac band touching point into two Dirac points (polarized in the linearly dispersing direction) by an amount that depends on the amplitude of the light. Motivated by recent pump-probe experiments, we investigated the non-equilibrium spectral properties and momentum-dependent spin-texture of our model in the Floquet state following a quench in absence of phonons, and in the presence of phonon dissipation that leads to a steady-state independent of the pump protocol.   Finally, we make connections to optical measurements by computing the frequency dependence of the longitudinal and transverse optical conductivity for this two-band model.  We analyze the various contributions from inter-band transitions and different Floquet modes. Our results suggest strategies for optically controlling band structures and experimentally measuring topological Floquet systems.
\end{abstract}
\maketitle
\section{Introduction}


Recent years have witnessed dramatic advances in understanding the topological properties of the band structure of quantum many-particle systems\cite{hasan_colloquium_2010, ando_topological_2013, qi_topological_2011, Moore2010}. These include time-reversal (TR) breaking integer quantum Hall systems, TR invariant two-dimensional quantum spin Hall systems, and three-dimensional topological insulators (TIs). When inter-particle interactions are included, the phenomenology is even more diverse\cite{Maciejko:np15, Stern:arcmp16, Krempa:arcm14, mesaros2013, Chen:prb13}. Certain isotropic low-energy dispersions are known to have particular stability conditions with respect to inter-particle interactions. For example, two-dimensional Dirac points are perturbatively stable to interactions, requiring a finite interaction strength to open a gap\cite{Meng:nat10, Hohenadler:prl11, Yu:prl11}, which underlies the low-energy properties of single-layer graphene\cite{Neto:rmp09}. By contrast, two-dimensional quadratic band touching points are known to be perturbatively unstable (i.e. a gap is opened, or the band touching point splits into two Dirac points) to interactions\cite{kaisun-prl103-2009}. 


On the other hand, anisotropic band touching points dominating the low-energy physics are more intriguing as both Coulomb interactions and disorder can have interesting consequences\cite{Adroguer2015}. Notably, semi-Dirac fermions have an anisotropic dispersion which displays a linear dispersion along one direction and a quadratic dispersion in the perpendicular direction\cite{Dietl2008, Banerjee2009}. Such a dispersion can be realized in phosphorene, in $\mathrm{TiO_2 / VO_2}$ superlattices\cite{Banerjee2009, Pardo2009, Pardo:prb10}, deformed graphene,  and $\mathrm{BEDT-TTF_2I_3}$ salt under pressure\cite{Kobayashi:prb11, Hasegawa:prb06, Yoshikazu:JPSJ13}. Systems with semi-Dirac band touching points are unstable to Coulomb interactions and display marginal Fermi liquid behavior with well-defined quasi-particles\cite{Zhao2016, Sriluckshmy2017, Cho2015}.


Another interesting class of topological states studied in recent years arises from the non-equilibrium generation of interesting band structures under the influence of a periodic drive\cite{Yan2017, Saha2016, Narayan2015, Du2017, Du2017a, Insulator2013, Fregoso2013}. At the non-interacting level, dramatic changes in the band structure can occur, including a change from a non-topological band structure to a topological one\cite{Kitagawa:prb10, Rudner:prx13, Katan:prl13, Lindner:prb13, Dora:prl12, Inoue:prl10, Cayssol:pss13, Kitagawa:prb11, Iadecola:prl13, Ezawa:prl13, Kemper:prb13, Rechtsman:nat13}. Two commonly discussed physical scenarios for periodically driven systems include periodic changes in the laser fields that establish the optical lattice potential for cold atom systems\cite{Jotzu:nat14, Bilitewski:pra2015} and solid state systems that are driven by a monochromatic laser field\cite{Fregoso2013, Sentef2014, Wang2013, Mahmood:nat16, Calvo:prb15, Dal:pra15, Perez:prb14, Perez:pra15}.  Recent work shows that a quadratic band touching point in two-dimensions has a gap opened by virtual two-photon absorption and emission processes in some cases,\cite{Du2017a} while it can be opened by one-photon processes in others.\cite{Du2017}  By contrast, linearly polarized light splits the quadratic band touching point into two Dirac points by an amount that depends only on the amplitude and polarization direction of the light\cite{Du2017a}. When inter-particle interactions are included, energy is typically absorbed from the periodic drive\cite{Hyungwon:pre14} and a closed many particle system will generically end up at infinite temperature in the infinite time limit, unless nongeneric conditions such as many body localization are present\cite{Ponte:prl15, Lazarides:prl15, Genske:pra15}. On the other hand, if the system is open, i.e. coupled to a bath such as phonons, it is possible for a balance?? to be established where the average energy (over a drive period) absorbed by the system from the drive can be released to the bath and a nonequilibrium steady state established\cite{Dehghani2014, Dehghani:prb16, Iadecola:prb15, Iadecola:prb15a, Seetharam:prx15, Shirai:pre15, Tsuji2009}. Previous studies have mostly been performed on Floquet steady states in systems with isotropic low-energy dispersions\cite{Insulator2013, Fregoso2013, Du2017a} while a thorough examination of anisotropic band touching points under periodic drive is still lacking.


In this paper, we focus on a periodically driven semi-Dirac band model on the honeycomb lattice. We demonstrate that circularly polarized light can induce a TR breaking topological band structure carrying finite Chern numbers in the non-equilibrium steady states, while linearly polarized light can split the semi-Dirac point into two linearly dispersing Dirac points.  A quench into the Floquet state yields a strongly momentum-dependent spin density. By contrast, we find an open semi-Dirac system with phonon dissipation can remove the anisotropy introduced by the quench from the initial state, which is qualitatively similar to the study of the Dirac dispersion\cite{Dehghani2014}.  We examine the spin-averaged ARPES spectrum, the time-averaged spin density, and we compute the longitudinal and Hall optical conductivity.  We analyze the contribution from different Floquet modes and emphasize the important differences between linearly polarized and circularly polarized driving fields. 


Our paper is organized as follows. In Sec.~\ref{sect:Lattice_Model_and_Band_Structure}, we describe
the lattice Hamiltonian we study, and in Secs.~\ref{sect:PERIODIC_DRIVE_UNDER_A_LASER_FIELD} and~\ref{sec:FLOQUET_THEORY} we discuss the influence of a monochromatic laser field of different polarizations, intensities, and frequencies on the Hamiltonian. In Sec.~\ref{sec:SPECTRAL_FUNCTION} we present the spectral function and time-averaged spin texture. In Sec.~\ref{sec:LONGITUDINAL_OPTICAL_CONDUCTIVITY}, we compute the finite-frequency longitudinal optical conductivity of the model for different laser parameters. In Sec.~\ref{sec:CHERN_NUMBER_AND_OPTICAL_HALL_CONDUCTIVITY} we address the Hall optical conductivity in comparison with the longitudinal components. In Sec.~\ref{sect:CONCLUSION_AND_DISCUSSION} we summarize the main conclusions of this work and discuss their relevance to real materials.  Details of the derivation of the longitudinal optical conductivity and the low energy effective model are presented in the Appendices.


\section{Lattice Model and Band Structure}
\label{sect:Lattice_Model_and_Band_Structure}
\begin{figure}[t]
\epsfig{file=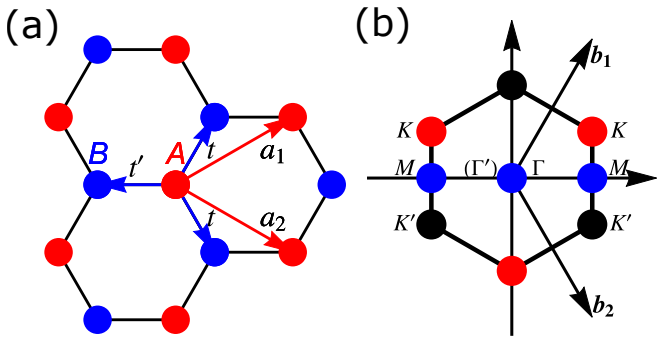,clip=0.1,width=\linewidth,angle=0}
\caption{(Color online) The honeycomb lattice and the first Brillouin zone. (a) The honeycomb lattice with primitive lattice vectors $a_1=a\left(\frac{3}{2},\frac{\sqrt{3}}{2}\right), a_2=a\left(\frac{3}{2},-\frac{\sqrt{3}}{2}\right)$. A, B sublattices are colored as red and blue respectively. (b) Reciprocal lattice vectors $b_1=2\pi/3a\left(1,\sqrt{3}\right), b_1=2\pi/3a\left(1,-\sqrt{3}\right)$ and the first Brillouin zone with high symmetry points are marked. We note that $\Gamma^{'}=b_1+b_2$ is equivalent to $\Gamma$ for the convenience of plotting band structures along $\Gamma \rightarrow M \rightarrow \Gamma^{'}$.}
\label{fig: honeycomb}
\end{figure}

\begin{figure}[t]
%
\epsfig{file=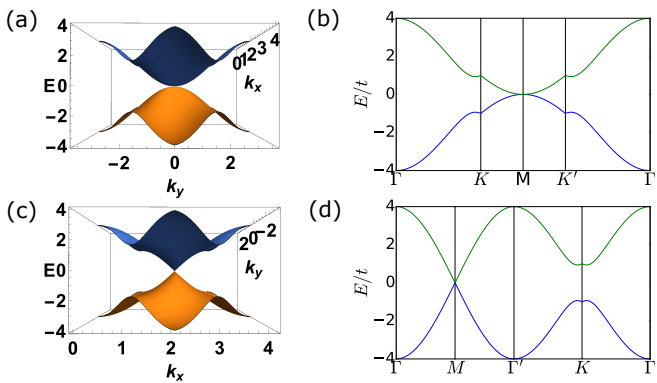,clip=0.1,width=\linewidth,angle=0}
\caption{(Color online) The band energies from Eq.(\ref{eq: dispersions}) with $t^{'}=2t$ and the semi-Dirac band touching point at $\mathbf{M}=(\frac{2 \pi}{3},0)$ for (a) Energy dispersions viewed along the $k_y$ axis, (b) dispersion along $\Gamma \rightarrow K \rightarrow M \rightarrow \Gamma$, (c) Energy dispersions viewed along the $k_x$ axis, (d) dispersion along $\Gamma \rightarrow M \rightarrow \Gamma^{'} \rightarrow K \rightarrow \Gamma$. }
\label{fig: dispersion}
\end{figure}

We study a honeycomb lattice model with anisotropic hopping that leads to semi-Dirac dispersions at low energy.  We also consider a coupling of electrons to a bath of phonons. The total Hamiltonian is
\begin{equation}
H=H_0+H_{ph}+H_c,
\label{eq: ham_tot}
\end{equation}
where $H_0$ is the tight-binding model with different values of nearest-neighbor (NN) hopping parameters that produces the semi-Dirac band touching point:
\begin{equation}
H_0=\sum _{\mathbf{l}} \left(t c_{B, \mathbf{l}+\mathbf{a}_1}^{\dagger }c_{A,\mathbf{l}}+t c_{B, \mathbf{l}+\mathbf{a}_2}^{\dagger }c_{A, \mathbf{l}}+t'c_{B, \mathbf{l}}^{\dagger
}c_{A, \mathbf{l}}\right)+h.c.,
\label{eq: ham_stat}
\end{equation}
and $H_{ph}$ is the phonon Hamiltonian, with $H_c$ the Hamiltonian describing the coupling of electrons and phonons.  In Fig. \ref{fig: honeycomb}, the primitive lattice vectors are chosen as (we set the lattice constant $a=1$ in the remainder of the paper),
\begin{equation}
\mathbf{a}_1=a\left(\frac{3}{2},\frac{\sqrt{3}}{2}\right), \mathbf{a}_2=a\left(\frac{3}{2},-\frac{\sqrt{3}}{2}\right),
\end{equation}
where $t$ is the NN hopping integral along $\delta_1=(1/2,\sqrt{3}/2)$ and $\delta_2=(1/2,-\sqrt{3}/2)$, $t'$ is the NN hopping integral along $\delta_3=(-1,0)$, $c_{A(B)i}, c^{\dagger}_{A(B)i}$ are creation and annihilation operators of electrons on the A(B) sublattices. The electron Hamiltonian $H_0$ can be Fourier transformed and then diagonalized. The electron dispersions and corresponding band structure are obtained from the eigenvalues:
\begin{widetext}
\begin{equation}
\epsilon _{\pm }(\mathbf{k})=\pm \sqrt{2t^2+t'^2+2t^2\cos  \sqrt{3}k_y+4t't \cos \left(\frac{3}{2}k_x\right)\cos \left(\frac{\sqrt{3}}{2}k_y\right)}.
\label{eq: dispersions}
\end{equation}
\end{widetext}
For \(t'\neq 2t\), there are two Dirac points in the first Brillouin zone. If we set \(t'=2t\), the dispersion
is quadratic along $k_y$ and linear along $k_x$ near the position of the band touching point
\begin{equation}
\mathbf{M}=\left(\frac{2\pi }{3a},0\right).
\end{equation}
The spectrum (Fig. \ref{fig: dispersion}) is linear in $k_x$ and quadratic in $k_y$. The standard $\mathbf{k}\cdot \mathbf{p}$ Hamiltonian reads
\begin{equation}
H_{SD}(\mathbf{k})=\frac{k_y^2}{2m}\sigma_x+v_F k_x \sigma_y,
\label{eq: semi-Dirac ham}
\end{equation}
with the effective mass $m=2/3t$ and the fermi velocity $v_F=3 t$. In the following sections, we set $t'=2t$ to investigate semi-Dirac points under the influence of a periodically driven electric field. Furthermore, dissipation from the environment affects the electron distribution and thus the spectral density together with the electrical transport coefficients. Here we consider dissipation due to coupling to two-dimensional phonons, similar to the approach of Refs. [\onlinecite{Dehghani2014, Dehghani2015, Dehghani2015a}]: the phonon part of Eq. (\ref{eq: ham_tot}) is a bilinear form of free boson operators:
\begin{equation}
H_{\mathrm{ph}}=\sum_{q,i=x,y}\omega_{q i} b^{\dagger}_{q i}b_{q i},
\end{equation}
and the electron-phonon coupling is specified as
\begin{equation}
H_{c}=\sum_{k q,\sigma, \sigma^{'}=A,B}\omega_{q i} c^{\dagger}_{k \sigma} \mathbf{A}_{ph} (q) \cdot \mathbf{\sigma}_{\sigma \sigma^{'}} c_{k \sigma^{'}},
\end{equation}
with
\begin{equation}
\mathbf{A}_{\mathrm{ph}} (q)=[\lambda_{x,q}(b^{\dagger}_{q x}+b_{-q x}), \lambda_{y,q}(b^{\dagger}_{q y}+b_{-q y})],
\end{equation}
representing the phonon field. Here, $\sigma, \sigma'=A, B$ are pseudo-spin labels of sublattices. Above we have made the assumption that phonon induced electron scattering with different quasi-momentum does not occur\cite{Dehghani2014, Dehghani2015, Dehghani2015a}. In the following calculations, the electronic states at different quasi-momenta $k$ are independently coupled to the reservoir and the broadening effect of electron-phonon interaction is not taken into account.

\section{PERIODIC DRIVE UNDER A LASER FIELD}
\label{sect:PERIODIC_DRIVE_UNDER_A_LASER_FIELD}

When the system is coupled to a laser field, the Hamiltonian is modified according to the Peierls substitution $\mathbf{k} \rightarrow \mathbf{k}+\mathbf{A}(t_1)$:
\begin{equation}
H_{\mathbf{k}}(t_1)=\sum _{\mathbf{k}} \left(c_{\mathbf{k} A}^{\dagger },c_{\mathbf{k} B}^{\dagger }\right)\left(
\begin{array}{cc}
 0 & h_{\mathbf{k}}^{A B}(t_1) \\
 \left[h_{\mathbf{k}}^{A B}(t_1)\right]^* & 0 \\
\end{array}
\right)\left(
\begin{array}{c}
 c_{\mathbf{k} A} \\
 c_{\mathbf{k} B} \\
\end{array}
\right),
\end{equation}
where we use $t_1$ as the time label to distinguish it from the hopping parameter and
\begin{equation}
h_{\mathbf{k}}^{A B}(t_1)=\sum _{i=1,2} t e^{i (\mathbf{k}+\mathbf{A}(t_1))\cdot \mathbf{\delta }_i}+t'e^{i (\mathbf{k}+\mathbf{A}(t_1))\cdot \mathbf{\delta }_3}.
\label{eq: ham_A}
\end{equation}
In Eq. (\ref{eq: ham_A}), we set Planck's constant $\hbar = 1$, the speed of light $c = 1$, and the charge of the electron $e = 1$, and adopt the Coulomb gauge by setting the scaler potential $\phi = 0$. We ignore the tiny effect of the magnetic field. The units of energy are
expressed in terms of the hopping $t$ and we set $t=1$. As \(h_{\mathbf{k}}^{A B}\) is not invariant under translation by integer multiples of \(n_ib_i\), we could recover the symmetry by the shift \(c_{\mathbf{k}\mathbf{
}B}\to c_{\mathbf{k}\mathbf{ }B}e^{i \mathbf{k}\mathbf{\cdot }\mathbf{\delta }_3}\):
\begin{equation}
h_{\mathbf{k}}^{A B}(t_1)=t'e^{i \mathbf{A}(t_1)\cdot \mathbf{\delta }_3}+\sum _{i=1,2} t e^{i \mathbf{k}\mathbf{\cdot }\mathbf{a}_i+i \mathbf{A}(t_1)\cdot \mathbf{\delta }_i}.
\end{equation}
Throughout this paper, circularly polarized laser
fields are expressed with the vector potential $\mathbf{A}(t_1)=A(\cos (\Omega  t_1),\sin (\Omega  t_1))$ and linear polarized laser fields are expressed with $\mathbf{A}(t_1)=A(\sin (\Omega  t_1), 0)$ and $\mathbf{A}(t_1)=A(0, \sin (\Omega  t_1))$ for the polarization along $k_x$ and $k_y$ direction, respectively, where $A$ is the amplitude and $\Omega$ the frequency of the laser.


\section{FLOQUET THEORY}
\label{sec:FLOQUET_THEORY}
\begin{figure}[ht]
%
%
\epsfig{file=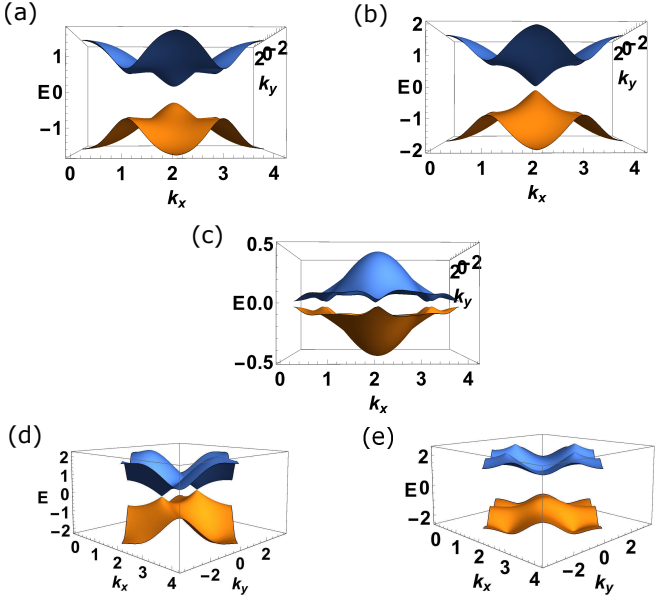,clip=0.1,width=\linewidth,angle=0}
\caption{(Color online) The Floquet band structures of the semi-Dirac honeycomb lattice model embedded in a normally incident polarized light for (a) circularly polarized light with $A=1.5, \Omega=5t$; (b) circularly polarized light with $A=1.5, \Omega=10t$; (c) circularly polarized light with $A=2.4, \Omega=5t$; (d) linearly polarized light with $A=1.5, \Omega=5t$ along $x$ direction; (e) linearly polarized light with $A=1.5, \Omega=5t$ along $y$ direction.  The semi-Dirac point linearly disperses in $k_x$ and has a quadratic dispersion along $k_y$. }
\label{fig: Floquet_bands}
\end{figure}

Since the laser field can be approximated as monochromatic (single frequency) light, it renders the Hamiltonian periodic in time: $H(t)=H(t+T)$, where $T$ is the period corresponding to $\Omega=2\pi/T$. In analogy to the periodicity in lattice translations that leads to Bloch's theorem, one can apply Floquet's theory\cite{GFloquet1883}. The Floquet eigenfunction can be expressed as
\begin{equation}
|\Psi_{k \alpha}(t)\rangle =e^{i \epsilon _{k \alpha }t}|\phi_{k \alpha }(t)\rangle,
\end{equation}
where $|\phi_{k \alpha }(t)\rangle = |\phi_{k \alpha }(t+T)\rangle$ are the Floquet quasimodes and $\epsilon_{k \alpha}$ is the corresponding quasienergy for band $\alpha$. Substituting this form of the wave function into the time-dependent Schrodinger equation, and defining the Floquet Hamiltonian operator as $\mathcal{H}(t) = H(t) - i\frac{\partial}{\partial t}$, one finds
\begin{equation}
\mathcal{H}(t)|\phi_{k \alpha }(t)\rangle = \epsilon_{k \alpha} |\phi_{k \alpha }(t)\rangle.
\end{equation}
By performing a Fourier transformation on time
\begin{eqnarray}
H_{\alpha \beta }^n &=& \frac{1}{T}\int _0^TH_{\alpha \beta }(t)\exp (-i n \Omega  t)dt, \nonumber \\
|\phi_{k \alpha }(t)\rangle &=& \sum_m e^{i m \Omega t} |\phi_{k \alpha }^m\rangle,
\label{eq: ham_n}
\end{eqnarray}
with $m,n=0, \pm 1, \pm 2,\ldots, \pm \infty$, one arrives at
\begin{equation}
\sum_m H_F^{n m} |\phi_{k \alpha }^m\rangle = \epsilon_{k \alpha} |\phi_{k \alpha }^n\rangle, 
\end{equation}
where
\begin{equation}
H_F^{n m}=H^{n-m}+n \Omega  I \delta _{n m},
\end{equation}
is the Floquet Hamiltonian living in the enlarged Floquet Hilbert space\cite{GFloquet1883}. In the lattice model we studied, 
\begin{widetext}
\begin{eqnarray}
h_{\mathbf{k}}^{A B}(m-n)&=&\frac{1}{T}\int _0^Th_{\mathbf{k}}^{A B}\left(t_1\right)\exp \left[-i (m-n) \Omega  t_1\right]dt_1 \nonumber \\
&=&\frac{1}{T}\int _0^Tdt_1\left[t e^{i \left[\frac{3k_x}{2}+\frac{\sqrt{3}k_y}{2}+\frac{A_x\left(t_1\right)}{2}+\frac{\sqrt{3}A_y\left(t_1\right)}{2}\right]}+t
e^{i \left[\frac{3k_x}{2}-\frac{\sqrt{3}k_y}{2}+\frac{A_x\left(t_1\right)}{2}-\frac{\sqrt{3}A_y\left(t_1\right)}{2}\right]}+t'e^{-i A_x\left(t_1\right)}\right] \nonumber \\
&& \times \exp \left[-i (m-n) \Omega  t_1\right].
\label{eq: ham_AB_mn}
\end{eqnarray}
\end{widetext}
In the numerical evaluation, we truncate the range of Floquet modes to $m,n=0,\pm1,\pm2,\pm3,\pm4$ and verified that a larger range of $m,n$ has little numerical impact on our results for the frequencies and electric field amplitudes we considered.

\subsection{Circularly polarized case}
For circularly polarized light, Eq. (\ref{eq: ham_AB_mn}) becomes
\begin{widetext}
\begin{equation}
h_{\mathbf{k}}^{A B}(m-n)=t e^{i \left[\frac{3k_x}{2}+\frac{\sqrt{3}k_y}{2}\right]}J_{m-n}(A)\exp \left[i (m-n) \frac{\pi }{6}\right]+t e^{i \left[\frac{3k_x}{2}-\frac{\sqrt{3}k_y}{2}\right]}J_{m-n}(A)\exp
\left[i (m-n) \frac{5\pi }{6}\right]+t'J_{m-n}(A)e^{-i (m-n)\frac{\pi }{2}},
\end{equation}
\end{widetext}
where $J_{n}(x)$ is the order-$n$ Bessel function of the first kind. The Floquet band structures are displayed in Fig. \ref{fig: Floquet_bands}(a-c). For $A=1.5, \Omega=5t$, there exists a gap between upper and lower band in a single Floquet copy. The gap size is $\Delta \approx 0.52 t$. As a comparison, the gap size for $A=1.5, \Omega=10t$ is $\Delta \approx 0.15 t$. They both hold finite Chern numbers $C=1$ for fully occupied ``lower" bands (which one is ``lower" is essentially a gauge choice; we refer here to the lower one in our figure), indicating the existence of topologically non-trivial transport properties under TR breaking circularly polarized light. A higher chern number with $C=2$ is realized by $A=2.4, \Omega=5t$ with a small gap size $\Delta \approx 0.05$. But unlike the graphene case\cite{Dehghani2015a}, we do not find $C=3$ for the semi-Dirac band structure in the presence of a laser field.

Moreover, we discovered that the leading order contribution to $\Delta$ is $O(\frac{A^4}{\Omega^2})$ in the small field amplitude and large frequency limit. This is revealed by the low energy effective theory in the high frequency expansion\cite{Bukov} up to $O(1/\Omega^2)$. The detailed analysis is given in Appendix \ref{sec:LOW_ENERGY_EFFECTIVE_HAMILTONIAN}. We would like to emphasize that this leading order contribution is different from either that of the quadratic band touching point\cite{Du2017a} $O(A^4/\Omega)$ or of the Dirac point\cite{Fregoso2013} $O(A^2/\Omega)$.

\subsection{Linearly polarized case}
In this work, we consider linear polarization in the $x$ and $y$ directions to reflect the symmetry of the semi-Dirac dispersion in our model. When the driven field is polarized in the $x$-direction, i.e., the linearly dispersing direction around the $M$ point according to Eq.~(\ref{eq: semi-Dirac ham}), Eq.~(\ref{eq: ham_AB_mn}) is reduced to
\begin{widetext}
\begin{equation}
h_{\mathbf{k}}^{A B}(m-n)=t e^{i \left[\frac{3k_x}{2}+\frac{\sqrt{3}k_y}{2}\right]}J_{m-n}\left(\frac{A}{2}\right)+t e^{i \left[\frac{3k_x}{2}-\frac{\sqrt{3}k_y}{2}\right]}J_{m-n}\left(\frac{A}{2}\right)+t'J_{n-m}(A).
\label{eq: hF_x}
\end{equation}
Similarly, if the polarization is in the $y$-direction, i.e., the quadratically dispersing direction around the $M$ point, Eq.~(\ref{eq: ham_AB_mn}) reads
\begin{equation}
h_{\mathbf{k}}^{A B}(m-n)=t e^{i \left[\frac{3k_x}{2}+\frac{\sqrt{3}k_y}{2}\right]}J_{m-n}\left(\frac{\sqrt{3}A}{2}\right)+t e^{i \left[\frac{3k_x}{2}-\frac{\sqrt{3}k_y}{2}\right]}J_{n-m}\left(\frac{\sqrt{3}A}{2}\right)+t'.
\end{equation}
\end{widetext}
Fig. \ref{fig: Floquet_bands}(c),(d) display the Floquet band structures under linearly polarized light along the $x$ and $y$ directions, respectively. Unlike circularly polarized light, linearly polarized light does not break the time-reversal symmetry\cite{Fregoso2013} and therefore the Chern number must be zero. For polarization along the quadratically dispersing direction, we find a gap opening induced at the band touching point. The gap size is of order ${\cal O}(A^2)$ and can be estimated from the zeroth order high frequency expansion of the low energy Hamiltonian in Appendix \ref{sec:LOW_ENERGY_EFFECTIVE_HAMILTONIAN}.  In contrast to the circular polarization case, the leading order contribution to the gap is independent of the driving frequency, $\Omega$. On the other hand, when the polarization is along the linearly dispersing direction, the bands remain gapless and the semi-Dirac band touching point described by Eq.~(\ref{eq: semi-Dirac ham}) is split into two single Dirac points. This particular feature of the Floquet bands can be roughly understood in the zeroth order high frequency expansion of the lattice model itself, which is the $n=0$ case of Eq.~(\ref{eq: hF_x}):
\begin{widetext}
\begin{equation}
h_{\mathbf{k}}^{A B}(m-n=0)=t e^{i \left[\frac{3k_x}{2}+\frac{\sqrt{3}k_y}{2}\right]}J_{0}\left(\frac{A}{2}\right)+t e^{i \left[\frac{3k_x}{2}-\frac{\sqrt{3}k_y}{2}\right]}J_{0}\left(\frac{A}{2}\right)+t'J_{0}(A).
\label{eq: hF_x_0}
\end{equation}
\end{widetext}
As long as $A \neq 0$, the coefficients in front of the phase factors in eq. (\ref{eq: hF_x_0}) do not change signs and the proportion $t/t^{'}$ is only renormalized by $J_0(A/2)/J_0(A)$, which leads to the splitting of semi-Dirac point into two Dirac points as $J_0(A/2)/J_0(A) \neq 1$ in analogy with different $t/t^{'}$ values in the static Hamiltonian Eq.~(\ref{eq: ham_stat}). Moreover, the two Dirac points are on the $k_y$ axis and their separation in the BZ is proportional to $A^2$ and independent of $\Omega$, up to leading order in the high frequency limit. This dependency is again captured by the low energy model (Appendix \ref{sec:LOW_ENERGY_EFFECTIVE_HAMILTONIAN}).

\section{SPECTRAL FUNCTION}
\label{sec:SPECTRAL_FUNCTION}
In this section, we examine the electronic spectral density of our model in both closed and open systems.  One can expand the fermionic operators in the quasimode basis at time \(t_0\),
\begin{equation}
c_{\mathbf{k}\sigma }\left(t_0\right)=\sum _{\alpha '} \phi _{\mathbf{k}\alpha '}^{\sigma }\left(t_0\right)\gamma _{\mathbf{k}\alpha '},
\end{equation}
where $\gamma _{\mathbf{k}\alpha '}$ annihilates a  particle in Floquet state $\mathbf{k}\alpha '$. In a closed system, the electron occupation probability is given by
\begin{equation}
\rho_{k, \alpha }=\left| \left\langle \phi _{k \alpha }(0)|\psi _{\text{in}, k}\right\rangle \right|^2,
\label{eq: rho_quench}
\end{equation}
where $|\psi _{\text{in}, k}\rangle$ is the initial state chosen to be the ground state of Eq.~(\ref{eq: ham_stat}). 

In an open system, we consider electrons coupled to a phonon bath described by Eq.~(\ref{eq: ham_tot}). We assume the reservoir of phonons to remain in thermal equilibrium at a temperature $T$. Inelastic scattering between electrons and phonons will cause the electron distribution function to relax and $\rho_{k, \alpha }$ can be solved using the methods of Ref.~[\onlinecite{Dehghani2014}].

The pseudo-spin-resolved ARPES spectrum is given by the lesser Greens function, \(i g_{\sigma \sigma }^<(k,\omega )\), with pseudospin labels \(\sigma =A,B\). The analytical expression is derived in Ref.~[\onlinecite{Dehghani2014}]:
\begin{eqnarray}
i g_{A A}^< (k,\omega)&=&2 \pi \sum _{m \alpha } \delta \left(\omega -\left[\epsilon_{k \alpha}-m \Omega \right]\right)\left|a_{m k \alpha}\right|
{}^2\rho _{k, \alpha}, \nonumber \\
i g_{B B}^<(k,\omega )&=&2\pi \sum _{m \alpha } \delta \left(\omega -\left[\epsilon _{k \alpha }-m \Omega \right]\right)\left|b_{m k \alpha }\right| {}^2\rho _{k, \alpha},\;
\end{eqnarray}
where \(a_{m k \alpha }\), \(b_{m k \alpha }\) are the Fourier transformed components of the Floquet eigenvectors,
\begin{equation}
|\phi _{k \alpha }(t)\rangle =\sum _{m\in \text{int}} e^{i m \Omega  t}\left(
\begin{array}{c}
 a_{m k \alpha } \\
 b_{m k \alpha } \\
\end{array}
\right).
\end{equation}
Then the total spectral density is \(A(k,\omega )=\text{Im}\left[\text{Tr}\left(g^R\right)\right]\).\cite{Dehghani2014} 

When the electron occupation probability is taken into account, the spectral density has an imbalance between upper and lower \(m=0\) Floquet bands. The total spectral density is a sum over psuedo-spin states \(i \sum _{\sigma
}g_{\sigma \sigma }^<(k,\omega )\).  It is also possible to measure the momentum resolved pseudo-spin polarization texture averaged over a period of the driving laser field, which is obtained from,
\begin{equation}
P_z\left(k_x,k_y\right)=i\int \frac{d\omega }{2\pi }\sum _{\sigma } \sigma  g_{\sigma \sigma }^<(k,\omega ).
\end{equation}
In the following, we will discuss the momentum and energy resolved spectral density and the momentum resolved pseudo-spin polarization for both closed and open systems under different polarizations of light.

\subsection{Circularly polarized light}
\begin{figure*}[t!]
\centering


\epsfig{file=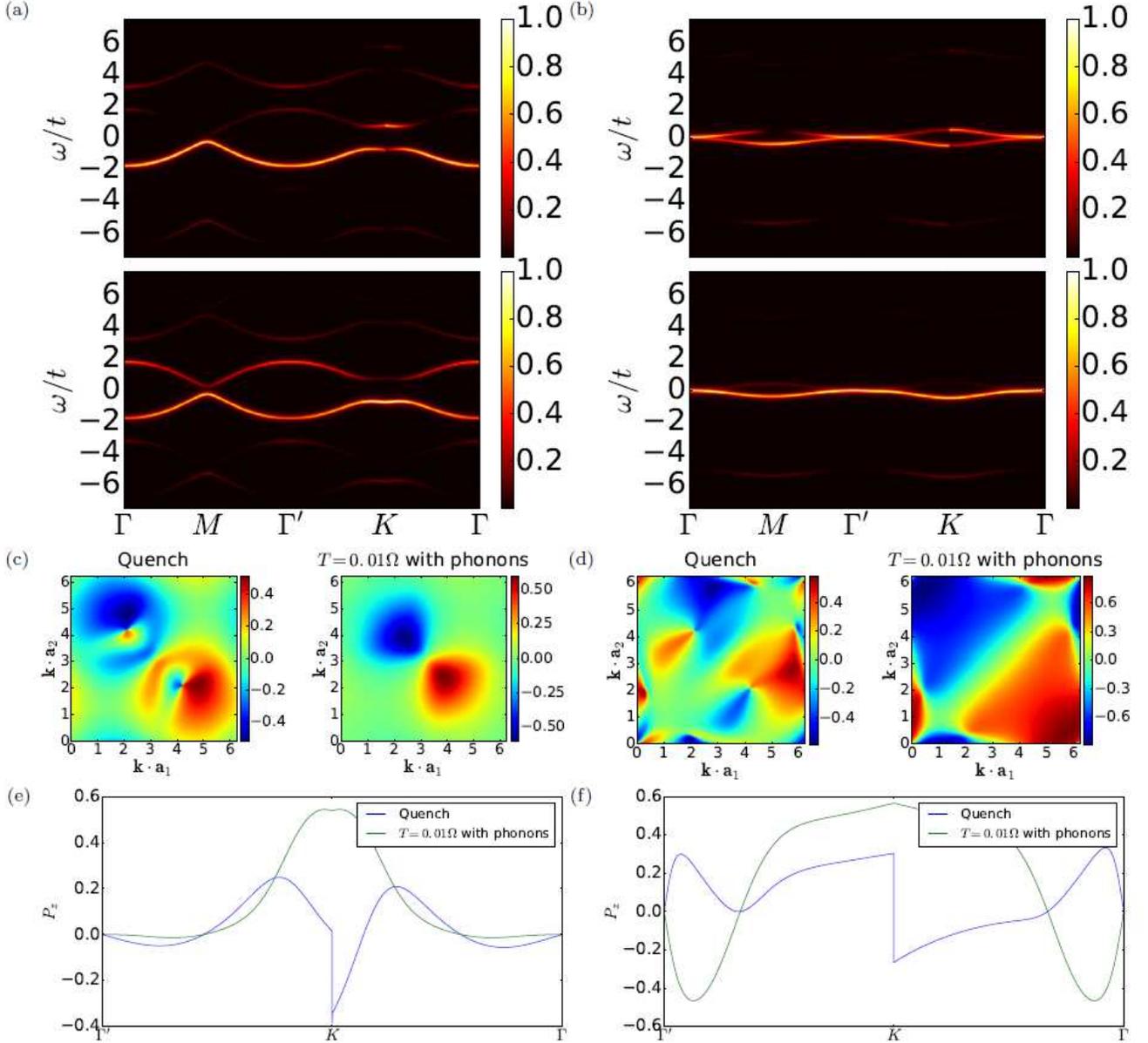,clip=0.1,width=\linewidth,angle=0}
\caption{(Color online) The spectral density of the semi-Dirac lattice model under the circularly polarized light. (a-b) ARPES spectrum \(i \sum _{\sigma
}g_{\sigma \sigma }^<(k,\omega )\) for (a) $A=1.5, \Omega=5t$ and (b) $A=2.4, \Omega=5t$ along high symmetry lines. Upper panel: without phonons and for a quench. Lower panel: steady state with phonons at $T=0.01 \Omega$; (c-d) Time averaged pseudo-spin density $P_z\left(k_x,k_y\right)$ for (c) $A=1.5, \Omega=5t$ and (d) $A=2.4, \Omega=5t$ in the first BZ with $\mathbf{k}\cdot \mathbf{a}_1$ and $\mathbf{k}\cdot \mathbf{a}_2$ as $x$ and $y$ axes. Left panel: without phonons and for a quench. Right panel: steady state with phonons at $T=0.01 \Omega$; (e-f) Time averaged pseudo-spin density $P_z\left(k_x,k_y\right)$ for (e) $A=1.5, \Omega=5t$ and (f) $A=2.4, \Omega=5t$ along the high symmetry line. In both (e) and (f), the pseudo-spin polarization textures show a discontinuity at the $K$ point following a quench.}
\label{fig: spectral_circ}
\end{figure*}

The ARPES spectrum and pseudo-spin textures in a circularly polarized laser field are shown in Fig. \ref{fig: spectral_circ}. From the spectral density along the high symmetry line, one can see the appearance of Floquet side bands. Without phonons, the system is quenched from its initial state to the Floquet eigenstate with an electron distribution density given by Eq.~(\ref{eq: rho_quench}). This is a highly nonthermal state in which the memory of the initial state is retained and the state does not thermalize\cite{Dehghani2014}. As a result, the ARPES spectrum intensity in Fig.\ref{fig: spectral_circ}(a-b) exhibits discontinuity at the $K$ point along $\Gamma \rightarrow K \rightarrow \Gamma^{'}$ and anisotropy at the $M$ point along $\Gamma \rightarrow M \rightarrow \Gamma^{'}$. The same character around $K$ can be observed in the momentum slices of $P_z\left(k_x,k_y\right)$ in Fig.\ref{fig: spectral_circ}(c-d). The asymmetry at $K$ and $M$ can be understood\cite{Dehghani2014}. When the initial gauge field is pointing along the $\hat{x}$ direction, $\rho^{quench}_{k \alpha}$ around $K$ and $M$ has a strong angle dependence on the phase angle $\theta(\mathbf{k})$ of the initial ground state. In the presence of a phonon bath, Fig.~\ref{fig: spectral_circ} shows that the lattice symmetry is retained in the ARPES spectrum and pseudo-spin textures, indicating that the phonons cause a loss of the memory of the initial states\cite{Dehghani2014} and lead to a nonequilibrium steady state distribution. In particular, the pseudo-spin texture with a phonon bath has perfect symmetry around $k_x$. For $A=1.5, \Omega=5t$, the band is predominantly of sublattice $B$ character\cite{Sentef2014} in the upper half of $k_x$-$k_y$ plane while sublattice $A$ dominates the lower half plane. The same phenomenon happens for the counterpart in $A=2.4, \Omega=5t$ except for the region near the BZ boundary, where positive and negative polarizations are separated by nodal lines. This is a strong indication of a further band inversion compared to $A=1.5, \Omega=5t$.

\subsection{Linearly polarized case}

\begin{figure*}[!t]

\epsfig{file=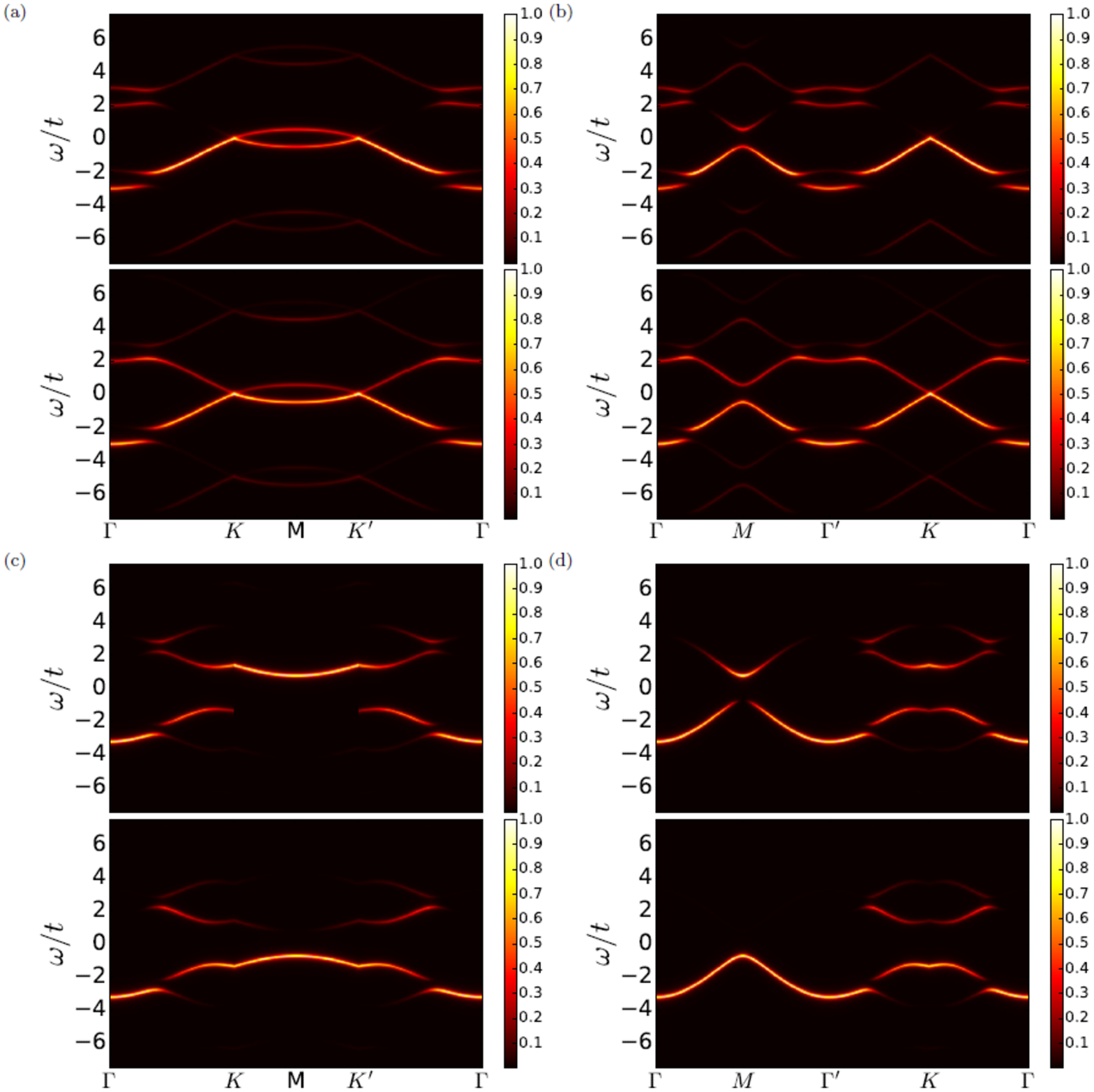,clip=0.1,width=\linewidth,angle=0}
\caption{(Color online) ARPES spectrum \(i \sum_{\sigma}g_{\sigma \sigma}^<(k,\omega)\) for the lattice model embedded in the linearly polarized light with $A=1.5, \Omega=5t$ along the high symmetry lines with (a-b) x-polarization, (c-d) y-polarization.  Upper panel: without phonons and for a quench. Lower panel: steady state with phonons at $T=0.01 \Omega$}
\label{fig: spectral_lin}
\end{figure*}

For comparison purposes, we plot the ARPES spectrum for our model in the presence of linearly polarized light in Fig. \ref{fig: spectral_lin}. Without phonons, it is noticeable that the asymmetry along $\Gamma \rightarrow K \rightarrow \Gamma^{'}$ and $\Gamma \rightarrow M \rightarrow \Gamma^{'}$ is no longer present for both $x$ and $y$ polarization in contrast with the circular polarization. This is due to the fact that at time $t_1=0$, the initial gauge field is exactly $0$ and the electron distribution is independent of the angle between the momentum and the gauge field around $K$ and $M$\cite{Dehghani2014}. In the presence of phonons, one can observe the spectral weight redistribution between upper and lower bands in all cases.

\section{LONGITUDINAL OPTICAL CONDUCTIVITY}
\label{sec:LONGITUDINAL_OPTICAL_CONDUCTIVITY}

Although angle resolved photoemission spectroscopy is a direct measurement of the energy spectrum in the system, it can only detect occupied states\cite{Freericks2009a}. Here we investigate the electromagnetic response of the system.\cite{Dehghani2015, Dehghani2015a}  In the following, we will present a thorough study of both the longitudinal and the Hall optical conductivity. In this section, our focus is on the longitudinal components of the ac conductivity, for which the formula is derived in Appendix~\ref{sec:DERIVATION_OF_LONGITUDINAL_OPTICAL_CONDUCTIVITY},
\begin{widetext}
\begin{eqnarray}
\text{Re}[\sigma _{i i}(\omega )]&=&\frac{1}{N}\sum _{\mathbf{k}} \sum _m D_{u i d}^m(\mathbf{k})D_{d i u}^{-m}(\mathbf{k})(\rho _{\mathbf{k}u}-\rho _{\mathbf{k}d}) \nonumber \\
& & \times\frac{-4\left(\epsilon _{\mathbf{k}d}-\epsilon _{\mathbf{k}u}-m
\Omega \right)\delta }{\left[\omega ^2-\left(\epsilon _{\mathbf{k}d}-\epsilon _{\mathbf{k}u}-m \Omega \right){}^2\right]{}^2+2\left(\omega ^2+\left(\epsilon
_{\mathbf{k}d}-\epsilon _{\mathbf{k}u}-m \Omega \right){}^2\right) \delta ^2},
\end{eqnarray}
\end{widetext}
where
\begin{equation}
D_{u i d}^m(\mathbf{k})=\sum _{n l} \langle \tilde{\phi }_{\mathbf{k}u}^n|[\frac{\partial h_{\mathbf{k}}^{m+n-l}}{\partial k_i}]|\tilde{\phi }_{\mathbf{k}d}^l\rangle.
\label{eq: long_cond}
\end{equation}
Eq.~(\ref{eq: long_cond}) can be seen as a generalization of the Kubo formula for a Floquet system. The total optical conductivity is comprised of contributions from different Floquet modes,
\begin{widetext}
\begin{eqnarray}
\text{Re} [\sigma_{i i}(\omega )]&=& \sum_m \text{Re} [\sigma^m_{i i}(\omega )], \nonumber \\
\text{Re} [\sigma^m_{i i}(\omega ]&=&\frac{1}{N}\sum _{\mathbf{k}} D_{u i d}^m(\mathbf{k})D_{d i u}^{-m}(\mathbf{k})(\rho _{\mathbf{k}u}-\rho _{\mathbf{k}d}) \nonumber \\
& & \times\frac{-4\left(\epsilon _{\mathbf{k}d}-\epsilon _{\mathbf{k}u}-m
\Omega \right)\delta }{\left[\omega ^2-\left(\epsilon _{\mathbf{k}d}-\epsilon _{\mathbf{k}u}-m \Omega \right){}^2\right]{}^2+2\left(\omega ^2+\left(\epsilon
_{\mathbf{k}d}-\epsilon _{\mathbf{k}u}-m \Omega \right){}^2\right) \delta ^2}.
\label{eq: long_cond_m}
\end{eqnarray}
\end{widetext}
It is worth pointing out that Eq.~(\ref{eq: long_cond_m}) has most of its weight coming from regions where $\omega \approx \left| \epsilon _{\mathbf{k}d}-\epsilon _{\mathbf{k}u}-m \Omega \right|$. In our study, $\sigma_{xx}$ is along the linearly dispersing direction while $\sigma_{yy}$ is along the quadratically dispersing direction.

\subsection{Circular Polarization}

\begin{figure*}[!t]

\epsfig{file=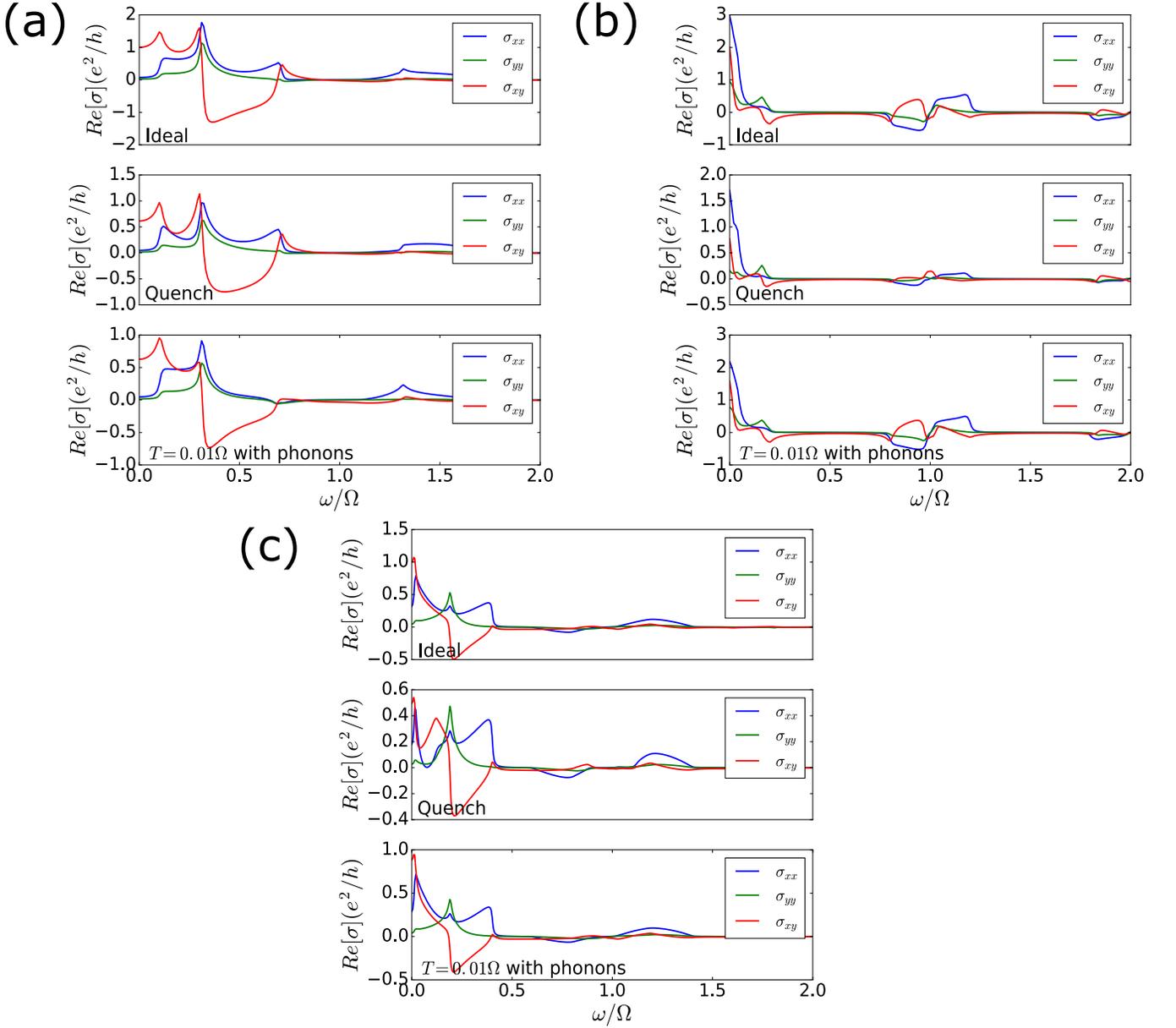,clip=0.1,width=\linewidth,angle=0}
\caption{(Color online) The longitudinal and transverse optical conductivity $\mathbf{Re}[\sigma]$ under a circularly polarized laser field as a function of the frequency of the probe light, in units of $e^2/h$. The driving laser frequency, laser amplitude, and the Chern number are (a) $\Omega=5t, A=1.5, C=1$, (b) $\Omega=5t, A=2.4, C=2$, (c) $\Omega=10t, A=1.5, C=1$. Top panel: ideal electron distribution with $\rho_{k d}-\rho_{k u}=1$. Middle panel: distribution following a quench. Bottom panel: steady state distribution with phonons at $T=0.01\Omega$}
\label{fig: long_cond_circ}
\end{figure*}
\begin{figure}[!t]
\epsfig{file=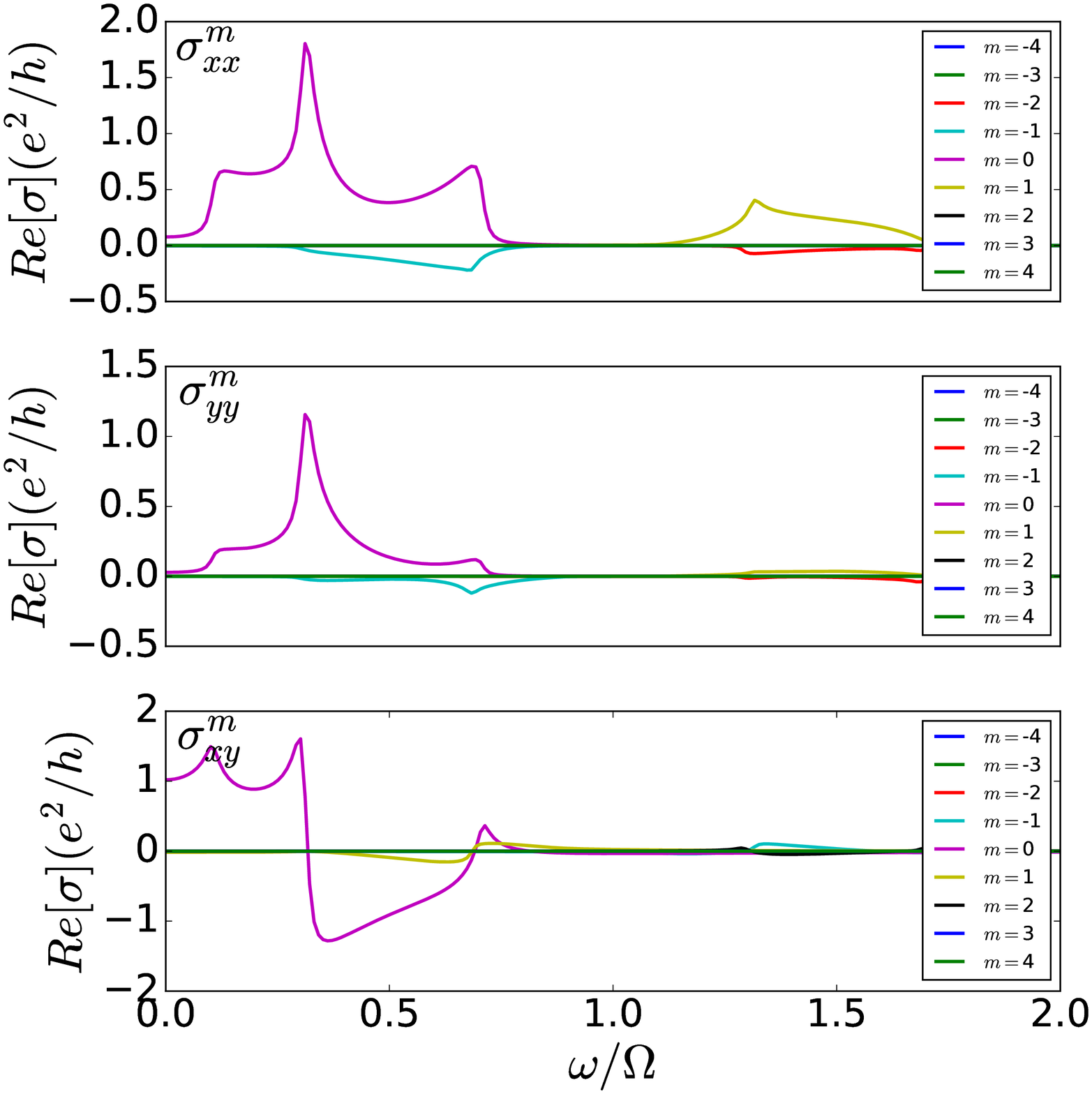,clip=0.1,width=\linewidth,angle=0}
\caption{(Color online) The longitudinal and transverse optical conductivity $\mathbf{Re}[\sigma]$ decomposed into different Floquet modes in the ideal case $\rho_{k d}-\rho_{k u}$ under a circularly polarized laser field as a function of the frequency of the probe light, in units of $e^2/h$ for $\Omega=5t, A=1.5, C=1$. Top panel: $\sigma^m_{xx}$. Middle panel: $\sigma^m_{yy}$. Bottom panel: $\sigma^m_{xy}$}
\label{fig: long_cond_m}
\end{figure}

From Fig.~\ref{fig: long_cond_circ}(a), one sees that $\sigma_{x x}$ and $\sigma_{y y}$ for $\Omega=5t, A=1.5$ are similar to each other in profile. Because of a finite gap in the Floquet band structure, $\sigma_{x x}$ and $\sigma_{y y}$ only have appreciable contributions from inter band quasi-electron excitations with $\omega \gtrsim \Delta$ (the gap).  Note that $\sigma_{x x}$ is larger than $\sigma_{y y}$ in the whole frequency range indicating a smaller effective mass generated by the laser field along the $x$-direction compared to the $y$-direction.  From Fig.~\ref{fig: long_cond_circ}(b), one sees that both $\sigma_{x x}$ and $\sigma_{y y}$ for $\Omega=5t, A=2.4$ become negative around $\omega \approx \Omega$. This is a characteristic feature of a Floquet system in the non-equilibrium steady state due to the non-zero electron distribution on the side bands. By examining Eq.~(\ref{eq: long_cond_m}), one can see that when $\omega \approx \left| \epsilon _{\mathbf{k}d}-\epsilon _{\mathbf{k}u}-m \Omega \right|$ for $m = -1$, the numerator can change sign if a quasi-electron can be excited from the lower band to the upper band by a single photon absorption. To illustrate this point, we plot Eq.~(\ref{eq: long_cond_m}) with all the Floquet modes for $A=1.5, \Omega=5t$ in the top and middle panel of Fig.~\ref{fig: long_cond_m}. We notice that the $m<0$ contributions are negative while the $m \geq 0$ contributions are all positive for both $\sigma^m_{xx}$ and $\sigma^m_{yy}$. Overall, the $m=0$ mode dominates the low frequency regime while $m \neq 0$ modes dominate the high frequency regime of the longitudinal optical conductivity. In Fig. \ref{fig: long_cond_circ}(c), we show the case of large driving frequency of the laser field: $A=1.5, \Omega=10t$, in which a sharp contrast between the profiles of $\sigma_{x x}$ and $\sigma_{y y}$ are observed. In the ideal case, the non-zero dc conductivity is due to the small gap size compared with the broadening parameter. In both closed and open systems, a finite electron distribution probability above the Fermi level will also lead to a finite contribution to the dc conductivity.

\subsection{Linear Polarization}

\begin{figure*}[!t]

\epsfig{file=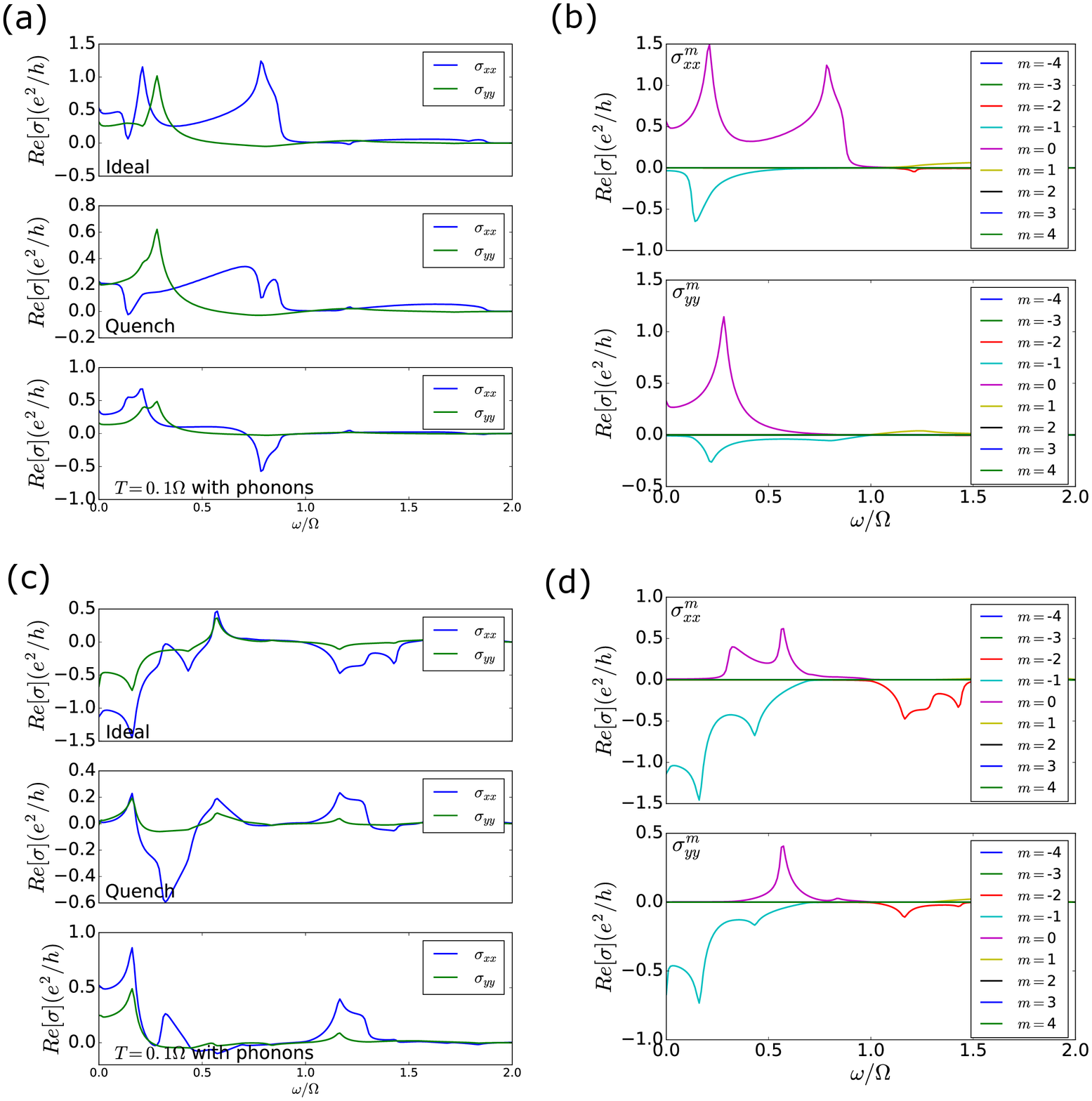,clip=0.1,width=\linewidth,angle=0}
\caption{(Color online) The longitudinal optical conductivity $\mathbf{Re}[\sigma]$ under a linearly polarized laser field and its decompositions into $\mathbf{Re}[\sigma^m]$ as a function of the frequency of the probe light with $\Omega=5t, A=1.5$. (a) $\mathbf{Re}[\sigma]$ for $x$-polarization, (b) $\mathbf{Re}[\sigma^m]$ for $x$-polarization, (c) $\mathbf{Re}[\sigma]$ for $y$-polarization, (d) $\mathbf{Re}[\sigma^m]$ for y polarization. Top panel: ideal electron distribution with $\rho_{k d}-\rho_{k u}=1$. Middle panel: distribution with quench. Bottom panel: distribution with phonons at $T=0.01\Omega$}
\label{fig: long_lin}
\end{figure*}




Next we turn to the optical conductivity of the linearly polarized driving field. Fig. \ref{fig: long_lin}(a) and (c) display $\sigma_{x x}$ and $\sigma_{y y}$ for polarization along $x$ and $y$ direction, respectively. It is obvious from both plots that $\sigma_{x x}$ and $\sigma_{y y}$ have significant difference in peak profile, indicating a sharp contrast between the gapless and gapped Floquet bands near the Fermi level. The shift in peak positions of the longitudinal optical conductivity for $x$ polarization (Fig. \ref{fig: long_lin}(a)) results from the anisotropy of the band structure along $x$ and $y$ directions, i.e. the splitting of the semi-Dirac point into single Dirac points separated along $k_y$. On the other hand, laser fields polarized along $y$-direction gives rise to a negative value for both $\sigma_{x x}$ and $\sigma_{y y}$ in the ideal case. This feature can be attributed to the gapless nature between the upper band of $m=0$ mode and the lower band of $m=1$ mode while $\epsilon_{k u}-\epsilon_{k d}$ holds a finite gap. To illustrate the point, we plot Eq.~(\ref{eq: long_cond_m}) with the ideal electron distribution in Fig. \ref{fig: long_lin}(b) and (d) corresponding to (a) and (c) respectively, where the dominant contribution at low probe frequency shifts from $\sigma^{m=0}$ to $\sigma^{m=-1}$ and changes sign by comparing (b) to (d). In both the quench and phonon panels of Fig. \ref{fig: long_lin}(c), the negative sign of $\sigma^{m=-1}$ is offset by the inversion in electron distribution between different Floquet modes and in consequence, the low frequency conductivity remains positive.

\section{CHERN NUMBER AND OPTICAL HALL CONDUCTIVITY}
\label{sec:CHERN_NUMBER_AND_OPTICAL_HALL_CONDUCTIVITY}

Starting from the linear response theory, the optical Hall conductivity is derived as\cite{Dehghani2015a}
\begin{widetext}
\begin{equation}
\sigma _{i j}(\omega )=-\frac{1}{N}\sum _{\mathbf{k},m} \left[\epsilon _{\mathbf{k}d}-\epsilon _{\mathbf{k}u}+m \Omega \right]^2F_{i j \mathbf{k}}^m\frac{\omega
^2-\left(\epsilon _{\mathbf{k}u}-\epsilon _{\mathbf{k}d}-m \Omega \right)^2-2i \omega  \delta }{\left[\omega ^2-\left(\epsilon _{\mathbf{k}u}-\epsilon _{\mathbf{k}d}-m
\Omega \right)^2\right]^2+4\omega ^2 \delta ^2}\langle \Psi \left(t_0\right)|[\gamma _{\mathbf{k}d}^{\dagger }\gamma _{\mathbf{k}d}-\gamma _{\mathbf{k}u}^{\dagger
}\gamma _{\mathbf{k}u}]|\Psi \left(t_0\right)\rangle,
\label{eq: hall_ac}
\end{equation}
where
\begin{equation}
F_{i j\mathbf{ }\mathbf{k}}^m=i\left[\sum _l \left\langle \tilde{\phi }_{\mathbf{k}u}^l|\partial _{k_i}\tilde{\phi }_{\mathbf{k}d}^{l-m}\right\rangle \sum _n
\left\langle \tilde{\phi }_{\mathbf{k}d}^n|\partial _{k_j}\tilde{\phi }_{\mathbf{k}u}^{n+m}\right\rangle -\sum _l \left\langle \tilde{\phi }_{\mathbf{k}d}^l|\partial
_{k_i}\tilde{\phi }_{\mathbf{k}u}^{l+m}\right\rangle \sum _n \left\langle \tilde{\phi }_{\mathbf{k}u}^n|\partial _{k_j}\tilde{\phi }_{\mathbf{k}d}^{n-m}\right\rangle
\right],
\end{equation}
is the Berry curvature and
\begin{equation}
A_{\beta  i \alpha }^m=\frac{1}{T}\int _0^Tdt e^{-i m \Omega  t}\left\langle \phi _{\mathbf{k}\beta }(t)|\partial _{k_i}\phi _{\mathbf{k}\alpha }(t)\right\rangle
\\
\\
=\frac{1}{T}\int _0^Tdt e^{-i m \Omega  t}\sum _l \sum _{l'} e^{i l' \Omega  t}e^{-i l \Omega  t}\left\langle \tilde{\phi }_{\mathbf{k}\beta }^l|\partial
_{k_i}\tilde{\phi }_{\mathbf{k}\alpha }^{l'}\right\rangle \\
\\
=\sum _l \left\langle \tilde{\phi }_{\mathbf{k}\beta }^l|\partial _{k_i}\tilde{\phi }_{\mathbf{k}\alpha }^{l+m}\right\rangle,
\end{equation}
\end{widetext}
is the Fourier transformed Berry connection. In the static limit $\omega \rightarrow 0$, Eq.(\ref{eq: hall_ac}), the dc Hall conductivity can be obtained as
\begin{equation}
\sigma _{i j}(\omega =0)=\int _{\text{BZ}}\frac{d^2k}{(2\pi )^2}\bar{F}_{\mathbf{k}d}\langle
\Psi \left(t_0\right)|[\gamma _{\mathbf{k}d}^{\dagger }\gamma _{\mathbf{k}d}-\gamma _{\mathbf{k}u}^{\dagger }\gamma _{\mathbf{k}u}]|\Psi \left(t_0\right)\rangle,
\label{eq: hall_dc}
\end{equation}
where
\begin{equation}
F_{\mathbf{k}d}(t)=i\left[\left\langle \partial _{k_i}\phi _{\mathbf{k}d}(t)|\partial _{k_j}\phi _{\mathbf{k}d}(t)\right\rangle -\left\langle \partial _{k_j}\phi
_{\mathbf{k}d}(t)|\partial _{k_i}\phi _{\mathbf{k}d}(t)\right\rangle \right],
\end{equation}
is the berry curvature in the real time. The above expression is in the unit of \(\frac{e^2}{\hbar }\), if we recover the units,
 \begin{equation}
\sigma _{i j}(\omega =0)=\frac{e^2}{h}\int _{\text{BZ}}\frac{d^2k}{(2\pi )^2}\bar{F}_{\mathbf{k}d}\langle
\Psi \left(t_0\right)|[\gamma _{\mathbf{k}d}^{\dagger }\gamma _{\mathbf{k}d}-\gamma _{\mathbf{k}u}^{\dagger }\gamma _{\mathbf{k}u}]|\Psi \left(t_0\right)\rangle.
\label{eq: hall_dc_units}
\end{equation}
In the ideal case, \(\langle \Psi \left(t_0\right)|[\gamma _{\mathbf{k}d}^{\dagger }\gamma _{\mathbf{k}d}-\gamma _{\mathbf{k}u}^{\dagger }\gamma _{\mathbf{k}u}]|\Psi
\left(t_0\right)\rangle =1\), Eq.(\ref{eq: hall_dc}) is reduced to
\begin{equation}
\sigma _{i j}(\omega =0)=\frac{e^2}{h}C,
\end{equation}
where $C$ is the Chern number computed as as
\begin{equation}
C=\frac{1}{2\pi }\int _{\text{BZ}}d^2k\bar{F}_{\mathbf{k}d}.
\end{equation}

In Fig. \ref{fig: long_cond_circ}, the Hall optical conductivity is plotted together with the longitudinal components for all cases we have examined in the system with circularly polarized laser fields. The main difference between the two is the oscillation between positive and negative values in $\sigma_{x y}$ for $\omega \ll \Omega$. In particular, for $\omega \approx max(\epsilon_{k u}-\epsilon_{k d})$, the optical Hall conductivity dips sharply into negative values while the longitudinal components are peaked due to the van Hove singularity. This can be explained by the different analytical behavior of the factors that include $\omega$ dependence. In Eq.(\ref{eq: long_cond_m}), the frequency dependent factor is sharply peaked at $\epsilon_{k d}-\epsilon_{k u}$ while the counterpart in Eq.(\ref{eq: hall_ac}) changes sign. In the bottom panel of Fig.~\ref{fig: long_cond_m}, we confirm that a sign change can happen within each $m$ in Eq.(\ref{eq: long_cond_m}).

\section{CONCLUSION AND DISCUSSION}
\label{sect:CONCLUSION_AND_DISCUSSION}

In this work, we addressed the influence of a laser driving field on a tight-binding model on the honeycomb lattice with a semi-Dirac dispersion at the low energies. We studied the effects of both circularly and linearly polarized light along two characteristic directions (reflecting the anisotropy of the semi-Dirac point) and analyzed different Floquet band structures from the low-energy effective Hamiltonian obtained in the high frequency limit. Compared to a nearest-neighbor hopping graphene model, the anisotropic band touching point we studied exhibits more diversity in gap openings, avoided crossings, and mixing between different Floquet side bands. We corroborated the richness by computing the ARPES spectrum and the pseudo-spin texture within quench scenario, and one that includes phonon dissipation.  These calculations connect with recent pump-probe experiments. In addition, we also studied the optical conductivity of the lattice model over the same conditions (quench and with phonons). The decomposition of the optical conductivity into different Floquet modes helps one better understand the Floquet band structure and connects to experiments by including realistic features of an electronic system in an open environment.

We would like to point out that the low energy Hamiltonian that captures the Floquet bands in our system is not the same as the semi-Dirac Hamiltonian with momentum replaced by Peierls substitution, 
\begin{equation}
H_{SD}(\mathbf{k},\mathbf{A})=\frac{(k_y+A_y)^2}{2m}\sigma_x+v_F (k_x+A_x) \sigma_y,
\label{eq: semi-Dirac ham(t)}
\end{equation}
which will only includes the vector potential $A_x(t)$ up to linear order. Thus, the gap size of leading order ${\cal O}(A^4/\Omega^2)$ is not captured correctly. Moreover, for a linearly polarized laser field applied along the $x$-direction, there is no splitting of the semi-Dirac point into two single Dirac-points along the $y$-direction in Eq.(\ref{eq: semi-Dirac ham(t)}). Our study highlights the fact that even though a leading order $\mathbf{k} \cdot \mathbf{p}$ Hamiltonian is a successful low-energy effective for the static Hamiltonian in equilibrium, its time-dependent counterpart by Peierls substitution can still hold different physical content from that of the correct low energy model.

Overall, our work broadens the scope for optically controlling band structures with topological band touching points and presents a detailed, experimentally accessible set of observables in lattice systems exposed to periodically driven laser field. The model we studied could be realized in modern cold atom experiments in optical lattices, in addition to the solid state systems we mentioned in the introduction. 

\section{Acknowledgment}   We thank Hsiang-Hsuan Hung, Chungwei Lin, Ming Xie, Allan H. MacDonald for helpful discussions. We gratefully acknowledge funding from ARO grant W911NF-14-1-0579, NSF DMR-1507621, and NSF MRSEC DMR-1720595. This work was performed in part at Aspen Center for Physics, which is supported by National Science Foundation grant PHY-1607611.


\bibliography{FloquetSemiDirac}

\begin{thebibliography}{69}%
\makeatletter
\providecommand \@ifxundefined [1]{%
 \@ifx{#1\undefined}
}%
\providecommand \@ifnum [1]{%
 \ifnum #1\expandafter \@firstoftwo
 \else \expandafter \@secondoftwo
 \fi
}%
\providecommand \@ifx [1]{%
 \ifx #1\expandafter \@firstoftwo
 \else \expandafter \@secondoftwo
 \fi
}%
\providecommand \natexlab [1]{#1}%
\providecommand \enquote  [1]{``#1''}%
\providecommand \bibnamefont  [1]{#1}%
\providecommand \bibfnamefont [1]{#1}%
\providecommand \citenamefont [1]{#1}%
\providecommand \href@noop [0]{\@secondoftwo}%
\providecommand \href [0]{\begingroup \@sanitize@url \@href}%
\providecommand \@href[1]{\@@startlink{#1}\@@href}%
\providecommand \@@href[1]{\endgroup#1\@@endlink}%
\providecommand \@sanitize@url [0]{\catcode `\\12\catcode `\$12\catcode
  `\&12\catcode `\#12\catcode `\^12\catcode `\_12\catcode `\%12\relax}%
\providecommand \@@startlink[1]{}%
\providecommand \@@endlink[0]{}%
\providecommand \url  [0]{\begingroup\@sanitize@url \@url }%
\providecommand \@url [1]{\endgroup\@href {#1}{\urlprefix }}%
\providecommand \urlprefix  [0]{URL }%
\providecommand \Eprint [0]{\href }%
\providecommand \doibase [0]{http://dx.doi.org/}%
\providecommand \selectlanguage [0]{\@gobble}%
\providecommand \bibinfo  [0]{\@secondoftwo}%
\providecommand \bibfield  [0]{\@secondoftwo}%
\providecommand \translation [1]{[#1]}%
\providecommand \BibitemOpen [0]{}%
\providecommand \bibitemStop [0]{}%
\providecommand \bibitemNoStop [0]{.\EOS\space}%
\providecommand \EOS [0]{\spacefactor3000\relax}%
\providecommand \BibitemShut  [1]{\csname bibitem#1\endcsname}%
\let\auto@bib@innerbib\@empty
\bibitem [{\citenamefont {Hasan}\ and\ \citenamefont
  {Kane}(2010)}]{hasan_colloquium_2010}%
  \BibitemOpen
  \bibfield  {author} {\bibinfo {author} {\bibfnamefont {M.~Z.}\ \bibnamefont
  {Hasan}}\ and\ \bibinfo {author} {\bibfnamefont {C.~L.}\ \bibnamefont
  {Kane}},\ }\href {\doibase 10.1103/RevModPhys.82.3045} {\bibfield  {journal}
  {\bibinfo  {journal} {Rev. Mod. Phys.}\ }\textbf {\bibinfo {volume} {82}},\
  \bibinfo {pages} {3045} (\bibinfo {year} {2010})}\BibitemShut {NoStop}%
\bibitem [{\citenamefont {Ando}(2013)}]{ando_topological_2013}%
  \BibitemOpen
  \bibfield  {author} {\bibinfo {author} {\bibfnamefont {Y.}~\bibnamefont
  {Ando}},\ }\href {\doibase 10.7566/JPSJ.82.102001} {\bibfield  {journal}
  {\bibinfo  {journal} {J. Phys. Soc. Jpn.}\ }\textbf {\bibinfo {volume}
  {82}},\ \bibinfo {pages} {102001} (\bibinfo {year} {2013})}\BibitemShut
  {NoStop}%
\bibitem [{\citenamefont {Qi}\ and\ \citenamefont
  {Zhang}(2011)}]{qi_topological_2011}%
  \BibitemOpen
  \bibfield  {author} {\bibinfo {author} {\bibfnamefont {X.-L.}\ \bibnamefont
  {Qi}}\ and\ \bibinfo {author} {\bibfnamefont {S.-C.}\ \bibnamefont {Zhang}},\
  }\href {\doibase 10.1103/RevModPhys.83.1057} {\bibfield  {journal} {\bibinfo
  {journal} {Rev. Mod. Phys.}\ }\textbf {\bibinfo {volume} {83}},\ \bibinfo
  {pages} {1057} (\bibinfo {year} {2011})}\BibitemShut {NoStop}%
\bibitem [{\citenamefont {Moore}(2010)}]{Moore2010}%
  \BibitemOpen
  \bibfield  {author} {\bibinfo {author} {\bibfnamefont {J.~E.}\ \bibnamefont
  {Moore}},\ }\href {\doibase 10.1038/nature08916} {\bibfield  {journal}
  {\bibinfo  {journal} {Nature}\ }\textbf {\bibinfo {volume} {464}},\ \bibinfo
  {pages} {194} (\bibinfo {year} {2010})}\BibitemShut {NoStop}%
\bibitem [{\citenamefont {Maciejko}\ and\ \citenamefont
  {Fiete}(2015)}]{Maciejko:np15}%
  \BibitemOpen
  \bibfield  {author} {\bibinfo {author} {\bibfnamefont {J.}~\bibnamefont
  {Maciejko}}\ and\ \bibinfo {author} {\bibfnamefont {G.~A.}\ \bibnamefont
  {Fiete}},\ }\href@noop {} {\bibfield  {journal} {\bibinfo  {journal} {Nat.
  Phys.}\ }\textbf {\bibinfo {volume} {11}},\ \bibinfo {pages} {385} (\bibinfo
  {year} {2015})}\BibitemShut {NoStop}%
\bibitem [{\citenamefont {Stern}(2016)}]{Stern:arcmp16}%
  \BibitemOpen
  \bibfield  {author} {\bibinfo {author} {\bibfnamefont {A.}~\bibnamefont
  {Stern}},\ }\href {\doibase 10.1146/annurev-conmatphys-031115-011559}
  {\bibfield  {journal} {\bibinfo  {journal} {Annual Review of Condensed Matter
  Physics}\ }\textbf {\bibinfo {volume} {7}},\ \bibinfo {pages} {349} (\bibinfo
  {year} {2016})}\BibitemShut {NoStop}%
\bibitem [{\citenamefont {Witczak-Krempa}\ \emph {et~al.}(2014)\citenamefont
  {Witczak-Krempa}, \citenamefont {Chen}, \citenamefont {Kim},\ and\
  \citenamefont {Balents}}]{Krempa:arcm14}%
  \BibitemOpen
  \bibfield  {author} {\bibinfo {author} {\bibfnamefont {W.}~\bibnamefont
  {Witczak-Krempa}}, \bibinfo {author} {\bibfnamefont {G.}~\bibnamefont
  {Chen}}, \bibinfo {author} {\bibfnamefont {Y.~B.}\ \bibnamefont {Kim}}, \
  and\ \bibinfo {author} {\bibfnamefont {L.}~\bibnamefont {Balents}},\
  }\href@noop {} {\bibfield  {journal} {\bibinfo  {journal} {Ann. Rev. Cond.
  Matt. Phys.}\ }\textbf {\bibinfo {volume} {5}},\ \bibinfo {pages} {57}
  (\bibinfo {year} {2014})}\BibitemShut {NoStop}%
\bibitem [{\citenamefont {Mesaros}\ and\ \citenamefont
  {Ran}(2013)}]{mesaros2013}%
  \BibitemOpen
  \bibfield  {author} {\bibinfo {author} {\bibfnamefont {A.}~\bibnamefont
  {Mesaros}}\ and\ \bibinfo {author} {\bibfnamefont {Y.}~\bibnamefont {Ran}},\
  }\href {\doibase 10.1103/PhysRevB.87.155115} {\bibfield  {journal} {\bibinfo
  {journal} {Phys. Rev. B}\ }\textbf {\bibinfo {volume} {87}},\ \bibinfo
  {pages} {155115} (\bibinfo {year} {2013})}\BibitemShut {NoStop}%
\bibitem [{\citenamefont {Chen}\ \emph {et~al.}(2013)\citenamefont {Chen},
  \citenamefont {Gu}, \citenamefont {Liu},\ and\ \citenamefont
  {Wen}}]{Chen:prb13}%
  \BibitemOpen
  \bibfield  {author} {\bibinfo {author} {\bibfnamefont {X.}~\bibnamefont
  {Chen}}, \bibinfo {author} {\bibfnamefont {Z.-C.}\ \bibnamefont {Gu}},
  \bibinfo {author} {\bibfnamefont {Z.-X.}\ \bibnamefont {Liu}}, \ and\
  \bibinfo {author} {\bibfnamefont {X.-G.}\ \bibnamefont {Wen}},\ }\href
  {\doibase 10.1103/PhysRevB.87.155114} {\bibfield  {journal} {\bibinfo
  {journal} {Phys. Rev. B}\ }\textbf {\bibinfo {volume} {87}},\ \bibinfo
  {pages} {155114} (\bibinfo {year} {2013})}\BibitemShut {NoStop}%
\bibitem [{\citenamefont {Meng}\ \emph {et~al.}(2010)\citenamefont {Meng},
  \citenamefont {Lang}, \citenamefont {Wessel}, \citenamefont {Assaad},\ and\
  \citenamefont {Muramatsu}}]{Meng:nat10}%
  \BibitemOpen
  \bibfield  {author} {\bibinfo {author} {\bibfnamefont {Z.~Y.}\ \bibnamefont
  {Meng}}, \bibinfo {author} {\bibfnamefont {T.~C.}\ \bibnamefont {Lang}},
  \bibinfo {author} {\bibfnamefont {S.}~\bibnamefont {Wessel}}, \bibinfo
  {author} {\bibfnamefont {F.~F.}\ \bibnamefont {Assaad}}, \ and\ \bibinfo
  {author} {\bibfnamefont {A.}~\bibnamefont {Muramatsu}},\ }\href@noop {}
  {\bibfield  {journal} {\bibinfo  {journal} {Nature}\ }\textbf {\bibinfo
  {volume} {464}},\ \bibinfo {pages} {847} (\bibinfo {year}
  {2010})}\BibitemShut {NoStop}%
\bibitem [{\citenamefont {Hohenadler}\ \emph {et~al.}(2011)\citenamefont
  {Hohenadler}, \citenamefont {Lang},\ and\ \citenamefont
  {Assaad}}]{Hohenadler:prl11}%
  \BibitemOpen
  \bibfield  {author} {\bibinfo {author} {\bibfnamefont {M.}~\bibnamefont
  {Hohenadler}}, \bibinfo {author} {\bibfnamefont {T.~C.}\ \bibnamefont
  {Lang}}, \ and\ \bibinfo {author} {\bibfnamefont {F.~F.}\ \bibnamefont
  {Assaad}},\ }\href {\doibase 10.1103/PhysRevLett.106.100403} {\bibfield
  {journal} {\bibinfo  {journal} {Phys. Rev. Lett.}\ }\textbf {\bibinfo
  {volume} {106}},\ \bibinfo {pages} {100403} (\bibinfo {year}
  {2011})}\BibitemShut {NoStop}%
\bibitem [{\citenamefont {Yu}\ \emph {et~al.}(2011)\citenamefont {Yu},
  \citenamefont {Xie},\ and\ \citenamefont {Li}}]{Yu:prl11}%
  \BibitemOpen
  \bibfield  {author} {\bibinfo {author} {\bibfnamefont {S.-L.}\ \bibnamefont
  {Yu}}, \bibinfo {author} {\bibfnamefont {X.~C.}\ \bibnamefont {Xie}}, \ and\
  \bibinfo {author} {\bibfnamefont {J.-X.}\ \bibnamefont {Li}},\ }\href
  {\doibase 10.1103/PhysRevLett.107.010401} {\bibfield  {journal} {\bibinfo
  {journal} {Phys. Rev. Lett.}\ }\textbf {\bibinfo {volume} {107}},\ \bibinfo
  {pages} {010401} (\bibinfo {year} {2011})}\BibitemShut {NoStop}%
\bibitem [{\citenamefont {Castro~Neto}\ \emph {et~al.}(2009)\citenamefont
  {Castro~Neto}, \citenamefont {Guinea}, \citenamefont {Peres}, \citenamefont
  {Novoselov},\ and\ \citenamefont {Geim}}]{Neto:rmp09}%
  \BibitemOpen
  \bibfield  {author} {\bibinfo {author} {\bibfnamefont {A.~H.}\ \bibnamefont
  {Castro~Neto}}, \bibinfo {author} {\bibfnamefont {F.}~\bibnamefont {Guinea}},
  \bibinfo {author} {\bibfnamefont {N.~M.~R.}\ \bibnamefont {Peres}}, \bibinfo
  {author} {\bibfnamefont {K.~S.}\ \bibnamefont {Novoselov}}, \ and\ \bibinfo
  {author} {\bibfnamefont {A.~K.}\ \bibnamefont {Geim}},\ }\href {\doibase
  10.1103/RevModPhys.81.109} {\bibfield  {journal} {\bibinfo  {journal} {Rev.
  Mod. Phys.}\ }\textbf {\bibinfo {volume} {81}},\ \bibinfo {pages} {109}
  (\bibinfo {year} {2009})}\BibitemShut {NoStop}%
\bibitem [{\citenamefont {Sun}\ \emph {et~al.}(2009)\citenamefont {Sun},
  \citenamefont {Yao}, \citenamefont {Fradkin},\ and\ \citenamefont
  {Kivelson}}]{kaisun-prl103-2009}%
  \BibitemOpen
  \bibfield  {author} {\bibinfo {author} {\bibfnamefont {K.}~\bibnamefont
  {Sun}}, \bibinfo {author} {\bibfnamefont {H.}~\bibnamefont {Yao}}, \bibinfo
  {author} {\bibfnamefont {E.}~\bibnamefont {Fradkin}}, \ and\ \bibinfo
  {author} {\bibfnamefont {S.~A.}\ \bibnamefont {Kivelson}},\ }\href {\doibase
  10.1103/PhysRevLett.103.046811} {\bibfield  {journal} {\bibinfo  {journal}
  {Phys. Rev. Lett.}\ }\textbf {\bibinfo {volume} {103}},\ \bibinfo {pages}
  {046811} (\bibinfo {year} {2009})}\BibitemShut {NoStop}%
\bibitem [{\citenamefont {Adroguer}\ \emph {et~al.}(2016)\citenamefont
  {Adroguer}, \citenamefont {Carpentier}, \citenamefont {Montambaux},\ and\
  \citenamefont {Orignac}}]{Adroguer2015}%
  \BibitemOpen
  \bibfield  {author} {\bibinfo {author} {\bibfnamefont {P.}~\bibnamefont
  {Adroguer}}, \bibinfo {author} {\bibfnamefont {D.}~\bibnamefont
  {Carpentier}}, \bibinfo {author} {\bibfnamefont {G.}~\bibnamefont
  {Montambaux}}, \ and\ \bibinfo {author} {\bibfnamefont {E.}~\bibnamefont
  {Orignac}},\ }\href {\doibase 10.1103/PhysRevB.93.125113} {\bibfield
  {journal} {\bibinfo  {journal} {Phys. Rev. B}\ }\textbf {\bibinfo {volume}
  {93}},\ \bibinfo {pages} {125113} (\bibinfo {year} {2016})}\BibitemShut
  {NoStop}%
\bibitem [{\citenamefont {Dietl}\ \emph {et~al.}(2008)\citenamefont {Dietl},
  \citenamefont {Pi{\'{e}}chon},\ and\ \citenamefont {Montambaux}}]{Dietl2008}%
  \BibitemOpen
  \bibfield  {author} {\bibinfo {author} {\bibfnamefont {P.}~\bibnamefont
  {Dietl}}, \bibinfo {author} {\bibfnamefont {F.}~\bibnamefont
  {Pi{\'{e}}chon}}, \ and\ \bibinfo {author} {\bibfnamefont {G.}~\bibnamefont
  {Montambaux}},\ }\href {\doibase 10.1103/PhysRevLett.100.236405} {\bibfield
  {journal} {\bibinfo  {journal} {Phys. Rev. Lett.}\ }\textbf {\bibinfo
  {volume} {100}},\ \bibinfo {pages} {1} (\bibinfo {year} {2008})}\BibitemShut
  {NoStop}%
\bibitem [{\citenamefont {Banerjee}\ \emph {et~al.}(2009)\citenamefont
  {Banerjee}, \citenamefont {Singh}, \citenamefont {Pardo},\ and\ \citenamefont
  {Pickett}}]{Banerjee2009}%
  \BibitemOpen
  \bibfield  {author} {\bibinfo {author} {\bibfnamefont {S.}~\bibnamefont
  {Banerjee}}, \bibinfo {author} {\bibfnamefont {R.~R.~P.}\ \bibnamefont
  {Singh}}, \bibinfo {author} {\bibfnamefont {V.}~\bibnamefont {Pardo}}, \ and\
  \bibinfo {author} {\bibfnamefont {W.~E.}\ \bibnamefont {Pickett}},\ }\href
  {\doibase 10.1103/PhysRevLett.103.016402} {\bibfield  {journal} {\bibinfo
  {journal} {Phys. Rev. Lett.}\ }\textbf {\bibinfo {volume} {103}},\ \bibinfo
  {pages} {15} (\bibinfo {year} {2009})}\BibitemShut {NoStop}%
\bibitem [{\citenamefont {Pardo}\ and\ \citenamefont
  {Pickett}(2009)}]{Pardo2009}%
  \BibitemOpen
  \bibfield  {author} {\bibinfo {author} {\bibfnamefont {V.}~\bibnamefont
  {Pardo}}\ and\ \bibinfo {author} {\bibfnamefont {W.~E.}\ \bibnamefont
  {Pickett}},\ }\href {\doibase 10.1103/PhysRevLett.102.166803} {\bibfield
  {journal} {\bibinfo  {journal} {Phys. Rev. Lett.}\ }\textbf {\bibinfo
  {volume} {102}},\ \bibinfo {pages} {2} (\bibinfo {year} {2009})}\BibitemShut
  {NoStop}%
\bibitem [{\citenamefont {Pardo}\ and\ \citenamefont
  {Pickett}(2010)}]{Pardo:prb10}%
  \BibitemOpen
  \bibfield  {author} {\bibinfo {author} {\bibfnamefont {V.}~\bibnamefont
  {Pardo}}\ and\ \bibinfo {author} {\bibfnamefont {W.~E.}\ \bibnamefont
  {Pickett}},\ }\href {\doibase 10.1103/PhysRevB.81.035111} {\bibfield
  {journal} {\bibinfo  {journal} {Phys. Rev. B}\ }\textbf {\bibinfo {volume}
  {81}},\ \bibinfo {pages} {035111} (\bibinfo {year} {2010})}\BibitemShut
  {NoStop}%
\bibitem [{\citenamefont {Kobayashi}\ \emph {et~al.}(2011)\citenamefont
  {Kobayashi}, \citenamefont {Suzumura}, \citenamefont {Pi\'echon},\ and\
  \citenamefont {Montambaux}}]{Kobayashi:prb11}%
  \BibitemOpen
  \bibfield  {author} {\bibinfo {author} {\bibfnamefont {A.}~\bibnamefont
  {Kobayashi}}, \bibinfo {author} {\bibfnamefont {Y.}~\bibnamefont {Suzumura}},
  \bibinfo {author} {\bibfnamefont {F.}~\bibnamefont {Pi\'echon}}, \ and\
  \bibinfo {author} {\bibfnamefont {G.}~\bibnamefont {Montambaux}},\ }\href
  {\doibase 10.1103/PhysRevB.84.075450} {\bibfield  {journal} {\bibinfo
  {journal} {Phys. Rev. B}\ }\textbf {\bibinfo {volume} {84}},\ \bibinfo
  {pages} {075450} (\bibinfo {year} {2011})}\BibitemShut {NoStop}%
\bibitem [{\citenamefont {Hasegawa}\ \emph {et~al.}(2006)\citenamefont
  {Hasegawa}, \citenamefont {Konno}, \citenamefont {Nakano},\ and\
  \citenamefont {Kohmoto}}]{Hasegawa:prb06}%
  \BibitemOpen
  \bibfield  {author} {\bibinfo {author} {\bibfnamefont {Y.}~\bibnamefont
  {Hasegawa}}, \bibinfo {author} {\bibfnamefont {R.}~\bibnamefont {Konno}},
  \bibinfo {author} {\bibfnamefont {H.}~\bibnamefont {Nakano}}, \ and\ \bibinfo
  {author} {\bibfnamefont {M.}~\bibnamefont {Kohmoto}},\ }\href {\doibase
  10.1103/PhysRevB.74.033413} {\bibfield  {journal} {\bibinfo  {journal} {Phys.
  Rev. B}\ }\textbf {\bibinfo {volume} {74}},\ \bibinfo {pages} {033413}
  (\bibinfo {year} {2006})}\BibitemShut {NoStop}%
\bibitem [{\citenamefont {Suzumura}\ \emph {et~al.}(2013)\citenamefont
  {Suzumura}, \citenamefont {Morinari},\ and\ \citenamefont
  {Piéchon}}]{Yoshikazu:JPSJ13}%
  \BibitemOpen
  \bibfield  {author} {\bibinfo {author} {\bibfnamefont {Y.}~\bibnamefont
  {Suzumura}}, \bibinfo {author} {\bibfnamefont {T.}~\bibnamefont {Morinari}},
  \ and\ \bibinfo {author} {\bibfnamefont {F.}~\bibnamefont {Piéchon}},\
  }\href {\doibase 10.7566/JPSJ.82.023708} {\bibfield  {journal} {\bibinfo
  {journal} {Journal of the Physical Society of Japan}\ }\textbf {\bibinfo
  {volume} {82}},\ \bibinfo {pages} {023708} (\bibinfo {year}
  {2013})}\BibitemShut {NoStop}%
\bibitem [{\citenamefont {Zhao}\ \emph {et~al.}(2016)\citenamefont {Zhao},
  \citenamefont {Wang}, \citenamefont {Wang},\ and\ \citenamefont
  {Liu}}]{Zhao2016}%
  \BibitemOpen
  \bibfield  {author} {\bibinfo {author} {\bibfnamefont {P.~L.}\ \bibnamefont
  {Zhao}}, \bibinfo {author} {\bibfnamefont {J.~R.}\ \bibnamefont {Wang}},
  \bibinfo {author} {\bibfnamefont {A.~M.}\ \bibnamefont {Wang}}, \ and\
  \bibinfo {author} {\bibfnamefont {G.~Z.}\ \bibnamefont {Liu}},\ }\href
  {\doibase 10.1103/PhysRevB.94.195114} {\bibfield  {journal} {\bibinfo
  {journal} {Phys. Rev. B}\ }\textbf {\bibinfo {volume} {94}},\ \bibinfo
  {pages} {1} (\bibinfo {year} {2016})}\BibitemShut {NoStop}%
\bibitem [{\citenamefont {Sriluckshmy}\ \emph {et~al.}(2017)\citenamefont
  {Sriluckshmy}, \citenamefont {Saha},\ and\ \citenamefont
  {Moessner}}]{Sriluckshmy2017}%
  \BibitemOpen
  \bibfield  {author} {\bibinfo {author} {\bibfnamefont {P.~V.}\ \bibnamefont
  {Sriluckshmy}}, \bibinfo {author} {\bibfnamefont {K.}~\bibnamefont {Saha}}, \
  and\ \bibinfo {author} {\bibfnamefont {R.}~\bibnamefont {Moessner}},\
  }\href@noop {} {\  (\bibinfo {year} {2017})},\ \Eprint
  {http://arxiv.org/abs/1709.00254v1} {arXiv:1709.00254v1} \BibitemShut
  {NoStop}%
\bibitem [{\citenamefont {Cho}\ and\ \citenamefont {Moon}(2015)}]{Cho2015}%
  \BibitemOpen
  \bibfield  {author} {\bibinfo {author} {\bibfnamefont {G.~Y.}\ \bibnamefont
  {Cho}}\ and\ \bibinfo {author} {\bibfnamefont {E.-G.}\ \bibnamefont {Moon}},\
  }\href {\doibase 10.1038/srep19198} {\bibfield  {journal} {\bibinfo
  {journal} {Sci. Rep.}\ }\textbf {\bibinfo {volume} {6}},\ \bibinfo {pages}
  {5} (\bibinfo {year} {2015})}\BibitemShut {NoStop}%
\bibitem [{\citenamefont {Yan}\ and\ \citenamefont {Wang}(2017)}]{Yan2017}%
  \BibitemOpen
  \bibfield  {author} {\bibinfo {author} {\bibfnamefont {Z.}~\bibnamefont
  {Yan}}\ and\ \bibinfo {author} {\bibfnamefont {Z.}~\bibnamefont {Wang}},\
  }\href {\doibase 10.1103/PhysRevB.96.041206} {\bibfield  {journal} {\bibinfo
  {journal} {Phys. Rev. B}\ }\textbf {\bibinfo {volume} {96}},\ \bibinfo
  {pages} {041206} (\bibinfo {year} {2017})}\BibitemShut {NoStop}%
\bibitem [{\citenamefont {Saha}(2016)}]{Saha2016}%
  \BibitemOpen
  \bibfield  {author} {\bibinfo {author} {\bibfnamefont {K.}~\bibnamefont
  {Saha}},\ }\href {\doibase 10.1103/PhysRevB.94.081103} {\bibfield  {journal}
  {\bibinfo  {journal} {Phys. Rev. B}\ }\textbf {\bibinfo {volume} {94}},\
  \bibinfo {pages} {1} (\bibinfo {year} {2016})}\BibitemShut {NoStop}%
\bibitem [{\citenamefont {Narayan}(2015)}]{Narayan2015}%
  \BibitemOpen
  \bibfield  {author} {\bibinfo {author} {\bibfnamefont {A.}~\bibnamefont
  {Narayan}},\ }\href {\doibase 10.1103/PhysRevB.91.205445} {\bibfield
  {journal} {\bibinfo  {journal} {Phys. Rev. B}\ }\textbf {\bibinfo {volume}
  {91}},\ \bibinfo {pages} {1} (\bibinfo {year} {2015})}\BibitemShut {NoStop}%
\bibitem [{\citenamefont {Du}\ and\ \citenamefont {Fiete}(2017)}]{Du2017}%
  \BibitemOpen
  \bibfield  {author} {\bibinfo {author} {\bibfnamefont {L.}~\bibnamefont
  {Du}}\ and\ \bibinfo {author} {\bibfnamefont {G.~A.}\ \bibnamefont {Fiete}},\
  }\href {\doibase 10.1103/PhysRevB.95.235309} {\bibfield  {journal} {\bibinfo
  {journal} {Phys. Rev. B}\ }\textbf {\bibinfo {volume} {95}},\ \bibinfo
  {pages} {235309} (\bibinfo {year} {2017})}\BibitemShut {NoStop}%
\bibitem [{\citenamefont {Du}\ \emph {et~al.}(2017)\citenamefont {Du},
  \citenamefont {Zhou},\ and\ \citenamefont {Fiete}}]{Du2017a}%
  \BibitemOpen
  \bibfield  {author} {\bibinfo {author} {\bibfnamefont {L.}~\bibnamefont
  {Du}}, \bibinfo {author} {\bibfnamefont {X.}~\bibnamefont {Zhou}}, \ and\
  \bibinfo {author} {\bibfnamefont {G.~A.}\ \bibnamefont {Fiete}},\ }\href
  {\doibase 10.1103/PhysRevB.95.035136} {\bibfield  {journal} {\bibinfo
  {journal} {Phys. Rev. B}\ }\textbf {\bibinfo {volume} {95}},\ \bibinfo
  {pages} {1} (\bibinfo {year} {2017})}\BibitemShut {NoStop}%
\bibitem [{\citenamefont {Wang}\ \emph {et~al.}(2013)\citenamefont {Wang},
  \citenamefont {Steinberg}, \citenamefont {Jarillo-Herrero},\ and\
  \citenamefont {Gedik}}]{Insulator2013}%
  \BibitemOpen
  \bibfield  {author} {\bibinfo {author} {\bibfnamefont {Y.~H.}\ \bibnamefont
  {Wang}}, \bibinfo {author} {\bibfnamefont {H.}~\bibnamefont {Steinberg}},
  \bibinfo {author} {\bibfnamefont {P.}~\bibnamefont {Jarillo-Herrero}}, \ and\
  \bibinfo {author} {\bibfnamefont {N.}~\bibnamefont {Gedik}},\ }\href
  {\doibase 10.1126/science.1239834} {\bibfield  {journal} {\bibinfo  {journal}
  {Science (80-. ).}\ }\textbf {\bibinfo {volume} {342}},\ \bibinfo {pages}
  {453} (\bibinfo {year} {2013})}\BibitemShut {NoStop}%
\bibitem [{\citenamefont {Fregoso}\ \emph {et~al.}(2013)\citenamefont
  {Fregoso}, \citenamefont {Wang}, \citenamefont {Gedik},\ and\ \citenamefont
  {Galitski}}]{Fregoso2013}%
  \BibitemOpen
  \bibfield  {author} {\bibinfo {author} {\bibfnamefont {B.~M.}\ \bibnamefont
  {Fregoso}}, \bibinfo {author} {\bibfnamefont {Y.~H.}\ \bibnamefont {Wang}},
  \bibinfo {author} {\bibfnamefont {N.}~\bibnamefont {Gedik}}, \ and\ \bibinfo
  {author} {\bibfnamefont {V.}~\bibnamefont {Galitski}},\ }\href {\doibase
  10.1103/PhysRevB.88.155129} {\bibfield  {journal} {\bibinfo  {journal} {Phys.
  Rev. B}\ }\textbf {\bibinfo {volume} {88}},\ \bibinfo {pages} {155129}
  (\bibinfo {year} {2013})}\BibitemShut {NoStop}%
\bibitem [{\citenamefont {Kitagawa}\ \emph {et~al.}(2010)\citenamefont
  {Kitagawa}, \citenamefont {Berg}, \citenamefont {Rudner},\ and\ \citenamefont
  {Demler}}]{Kitagawa:prb10}%
  \BibitemOpen
  \bibfield  {author} {\bibinfo {author} {\bibfnamefont {T.}~\bibnamefont
  {Kitagawa}}, \bibinfo {author} {\bibfnamefont {E.}~\bibnamefont {Berg}},
  \bibinfo {author} {\bibfnamefont {M.}~\bibnamefont {Rudner}}, \ and\ \bibinfo
  {author} {\bibfnamefont {E.}~\bibnamefont {Demler}},\ }\href {\doibase
  10.1103/PhysRevB.82.235114} {\bibfield  {journal} {\bibinfo  {journal} {Phys.
  Rev. B}\ }\textbf {\bibinfo {volume} {82}},\ \bibinfo {pages} {235114}
  (\bibinfo {year} {2010})}\BibitemShut {NoStop}%
\bibitem [{\citenamefont {Rudner}\ \emph {et~al.}(2013)\citenamefont {Rudner},
  \citenamefont {Lindner}, \citenamefont {Berg},\ and\ \citenamefont
  {Levin}}]{Rudner:prx13}%
  \BibitemOpen
  \bibfield  {author} {\bibinfo {author} {\bibfnamefont {M.~S.}\ \bibnamefont
  {Rudner}}, \bibinfo {author} {\bibfnamefont {N.~H.}\ \bibnamefont {Lindner}},
  \bibinfo {author} {\bibfnamefont {E.}~\bibnamefont {Berg}}, \ and\ \bibinfo
  {author} {\bibfnamefont {M.}~\bibnamefont {Levin}},\ }\href {\doibase
  10.1103/PhysRevX.3.031005} {\bibfield  {journal} {\bibinfo  {journal} {Phys.
  Rev. X}\ }\textbf {\bibinfo {volume} {3}},\ \bibinfo {pages} {031005}
  (\bibinfo {year} {2013})}\BibitemShut {NoStop}%
\bibitem [{\citenamefont {Katan}\ and\ \citenamefont
  {Podolsky}(2013)}]{Katan:prl13}%
  \BibitemOpen
  \bibfield  {author} {\bibinfo {author} {\bibfnamefont {Y.~T.}\ \bibnamefont
  {Katan}}\ and\ \bibinfo {author} {\bibfnamefont {D.}~\bibnamefont
  {Podolsky}},\ }\href {\doibase 10.1103/PhysRevLett.110.016802} {\bibfield
  {journal} {\bibinfo  {journal} {Phys. Rev. Lett.}\ }\textbf {\bibinfo
  {volume} {110}},\ \bibinfo {pages} {016802} (\bibinfo {year}
  {2013})}\BibitemShut {NoStop}%
\bibitem [{\citenamefont {Lindner}\ \emph {et~al.}(2013)\citenamefont
  {Lindner}, \citenamefont {Bergman}, \citenamefont {Refael},\ and\
  \citenamefont {Galitski}}]{Lindner:prb13}%
  \BibitemOpen
  \bibfield  {author} {\bibinfo {author} {\bibfnamefont {N.~H.}\ \bibnamefont
  {Lindner}}, \bibinfo {author} {\bibfnamefont {D.~L.}\ \bibnamefont
  {Bergman}}, \bibinfo {author} {\bibfnamefont {G.}~\bibnamefont {Refael}}, \
  and\ \bibinfo {author} {\bibfnamefont {V.}~\bibnamefont {Galitski}},\ }\href
  {\doibase 10.1103/PhysRevB.87.235131} {\bibfield  {journal} {\bibinfo
  {journal} {Phys. Rev. B}\ }\textbf {\bibinfo {volume} {87}},\ \bibinfo
  {pages} {235131} (\bibinfo {year} {2013})}\BibitemShut {NoStop}%
\bibitem [{\citenamefont {D\'ora}\ \emph {et~al.}(2012)\citenamefont {D\'ora},
  \citenamefont {Cayssol}, \citenamefont {Simon},\ and\ \citenamefont
  {Moessner}}]{Dora:prl12}%
  \BibitemOpen
  \bibfield  {author} {\bibinfo {author} {\bibfnamefont {B.}~\bibnamefont
  {D\'ora}}, \bibinfo {author} {\bibfnamefont {J.}~\bibnamefont {Cayssol}},
  \bibinfo {author} {\bibfnamefont {F.}~\bibnamefont {Simon}}, \ and\ \bibinfo
  {author} {\bibfnamefont {R.}~\bibnamefont {Moessner}},\ }\href {\doibase
  10.1103/PhysRevLett.108.056602} {\bibfield  {journal} {\bibinfo  {journal}
  {Phys. Rev. Lett.}\ }\textbf {\bibinfo {volume} {108}},\ \bibinfo {pages}
  {056602} (\bibinfo {year} {2012})}\BibitemShut {NoStop}%
\bibitem [{\citenamefont {Inoue}\ and\ \citenamefont
  {Tanaka}(2010)}]{Inoue:prl10}%
  \BibitemOpen
  \bibfield  {author} {\bibinfo {author} {\bibfnamefont {J.-i.}\ \bibnamefont
  {Inoue}}\ and\ \bibinfo {author} {\bibfnamefont {A.}~\bibnamefont {Tanaka}},\
  }\href {\doibase 10.1103/PhysRevLett.105.017401} {\bibfield  {journal}
  {\bibinfo  {journal} {Phys. Rev. Lett.}\ }\textbf {\bibinfo {volume} {105}},\
  \bibinfo {pages} {017401} (\bibinfo {year} {2010})}\BibitemShut {NoStop}%
\bibitem [{\citenamefont {Cayssol}\ \emph {et~al.}(2013)\citenamefont
  {Cayssol}, \citenamefont {Dora}, \citenamefont {Simon},\ and\ \citenamefont
  {Moessner}}]{Cayssol:pss13}%
  \BibitemOpen
  \bibfield  {author} {\bibinfo {author} {\bibfnamefont {J.}~\bibnamefont
  {Cayssol}}, \bibinfo {author} {\bibfnamefont {B.}~\bibnamefont {Dora}},
  \bibinfo {author} {\bibfnamefont {F.}~\bibnamefont {Simon}}, \ and\ \bibinfo
  {author} {\bibfnamefont {R.}~\bibnamefont {Moessner}},\ }\href {\doibase
  10.1002/pssr.201206451} {\bibfield  {journal} {\bibinfo  {journal} {physica
  status solidi (RRL) ‚Rapid Research Letters}\ }\textbf {\bibinfo {volume}
  {7}},\ \bibinfo {pages} {101} (\bibinfo {year} {2013})}\BibitemShut {NoStop}%
\bibitem [{\citenamefont {Kitagawa}\ \emph {et~al.}(2011)\citenamefont
  {Kitagawa}, \citenamefont {Oka}, \citenamefont {Brataas}, \citenamefont
  {Fu},\ and\ \citenamefont {Demler}}]{Kitagawa:prb11}%
  \BibitemOpen
  \bibfield  {author} {\bibinfo {author} {\bibfnamefont {T.}~\bibnamefont
  {Kitagawa}}, \bibinfo {author} {\bibfnamefont {T.}~\bibnamefont {Oka}},
  \bibinfo {author} {\bibfnamefont {A.}~\bibnamefont {Brataas}}, \bibinfo
  {author} {\bibfnamefont {L.}~\bibnamefont {Fu}}, \ and\ \bibinfo {author}
  {\bibfnamefont {E.}~\bibnamefont {Demler}},\ }\href {\doibase
  10.1103/PhysRevB.84.235108} {\bibfield  {journal} {\bibinfo  {journal} {Phys.
  Rev. B}\ }\textbf {\bibinfo {volume} {84}},\ \bibinfo {pages} {235108}
  (\bibinfo {year} {2011})}\BibitemShut {NoStop}%
\bibitem [{\citenamefont {Iadecola}\ \emph {et~al.}(2013)\citenamefont
  {Iadecola}, \citenamefont {Campbell}, \citenamefont {Chamon}, \citenamefont
  {Hou}, \citenamefont {Jackiw}, \citenamefont {Pi},\ and\ \citenamefont
  {Kusminskiy}}]{Iadecola:prl13}%
  \BibitemOpen
  \bibfield  {author} {\bibinfo {author} {\bibfnamefont {T.}~\bibnamefont
  {Iadecola}}, \bibinfo {author} {\bibfnamefont {D.}~\bibnamefont {Campbell}},
  \bibinfo {author} {\bibfnamefont {C.}~\bibnamefont {Chamon}}, \bibinfo
  {author} {\bibfnamefont {C.-Y.}\ \bibnamefont {Hou}}, \bibinfo {author}
  {\bibfnamefont {R.}~\bibnamefont {Jackiw}}, \bibinfo {author} {\bibfnamefont
  {S.-Y.}\ \bibnamefont {Pi}}, \ and\ \bibinfo {author} {\bibfnamefont {S.~V.}\
  \bibnamefont {Kusminskiy}},\ }\href {\doibase 10.1103/PhysRevLett.110.176603}
  {\bibfield  {journal} {\bibinfo  {journal} {Phys. Rev. Lett.}\ }\textbf
  {\bibinfo {volume} {110}},\ \bibinfo {pages} {176603} (\bibinfo {year}
  {2013})}\BibitemShut {NoStop}%
\bibitem [{\citenamefont {Ezawa}(2013)}]{Ezawa:prl13}%
  \BibitemOpen
  \bibfield  {author} {\bibinfo {author} {\bibfnamefont {M.}~\bibnamefont
  {Ezawa}},\ }\href {\doibase 10.1103/PhysRevLett.110.026603} {\bibfield
  {journal} {\bibinfo  {journal} {Phys. Rev. Lett.}\ }\textbf {\bibinfo
  {volume} {110}},\ \bibinfo {pages} {026603} (\bibinfo {year}
  {2013})}\BibitemShut {NoStop}%
\bibitem [{\citenamefont {Kemper}\ \emph {et~al.}(2013)\citenamefont {Kemper},
  \citenamefont {Sentef}, \citenamefont {Moritz}, \citenamefont {Kao},
  \citenamefont {Shen}, \citenamefont {Freericks},\ and\ \citenamefont
  {Devereaux}}]{Kemper:prb13}%
  \BibitemOpen
  \bibfield  {author} {\bibinfo {author} {\bibfnamefont {A.~F.}\ \bibnamefont
  {Kemper}}, \bibinfo {author} {\bibfnamefont {M.}~\bibnamefont {Sentef}},
  \bibinfo {author} {\bibfnamefont {B.}~\bibnamefont {Moritz}}, \bibinfo
  {author} {\bibfnamefont {C.~C.}\ \bibnamefont {Kao}}, \bibinfo {author}
  {\bibfnamefont {Z.~X.}\ \bibnamefont {Shen}}, \bibinfo {author}
  {\bibfnamefont {J.~K.}\ \bibnamefont {Freericks}}, \ and\ \bibinfo {author}
  {\bibfnamefont {T.~P.}\ \bibnamefont {Devereaux}},\ }\href {\doibase
  10.1103/PhysRevB.87.235139} {\bibfield  {journal} {\bibinfo  {journal} {Phys.
  Rev. B}\ }\textbf {\bibinfo {volume} {87}},\ \bibinfo {pages} {235139}
  (\bibinfo {year} {2013})}\BibitemShut {NoStop}%
\bibitem [{\citenamefont {Rechtsman}\ \emph {et~al.}(2013)\citenamefont
  {Rechtsman}, \citenamefont {Zeuner}, \citenamefont {Plotnik}, \citenamefont
  {Lumer}, \citenamefont {Podolsky}, \citenamefont {Dreisow}, \citenamefont
  {Nolte}, \citenamefont {Segev},\ and\ \citenamefont
  {Szameit}}]{Rechtsman:nat13}%
  \BibitemOpen
  \bibfield  {author} {\bibinfo {author} {\bibfnamefont {M.~C.}\ \bibnamefont
  {Rechtsman}}, \bibinfo {author} {\bibfnamefont {J.~M.}\ \bibnamefont
  {Zeuner}}, \bibinfo {author} {\bibfnamefont {Y.}~\bibnamefont {Plotnik}},
  \bibinfo {author} {\bibfnamefont {Y.}~\bibnamefont {Lumer}}, \bibinfo
  {author} {\bibfnamefont {D.}~\bibnamefont {Podolsky}}, \bibinfo {author}
  {\bibfnamefont {F.}~\bibnamefont {Dreisow}}, \bibinfo {author} {\bibfnamefont
  {S.}~\bibnamefont {Nolte}}, \bibinfo {author} {\bibfnamefont
  {M.}~\bibnamefont {Segev}}, \ and\ \bibinfo {author} {\bibfnamefont
  {A.}~\bibnamefont {Szameit}},\ }\href {\doibase 10.1038/nature12066}
  {\bibfield  {journal} {\bibinfo  {journal} {Nature}\ }\textbf {\bibinfo
  {volume} {496}},\ \bibinfo {pages} {196} (\bibinfo {year}
  {2013})}\BibitemShut {NoStop}%
\bibitem [{\citenamefont {Jotzu}\ \emph {et~al.}(2014)\citenamefont {Jotzu},
  \citenamefont {Messer}, \citenamefont {Desbuquois}, \citenamefont {Lebrat},
  \citenamefont {Uehlinger}, \citenamefont {Greif},\ and\ \citenamefont
  {Esslinger}}]{Jotzu:nat14}%
  \BibitemOpen
  \bibfield  {author} {\bibinfo {author} {\bibfnamefont {G.}~\bibnamefont
  {Jotzu}}, \bibinfo {author} {\bibfnamefont {M.}~\bibnamefont {Messer}},
  \bibinfo {author} {\bibfnamefont {R.}~\bibnamefont {Desbuquois}}, \bibinfo
  {author} {\bibfnamefont {M.}~\bibnamefont {Lebrat}}, \bibinfo {author}
  {\bibfnamefont {T.}~\bibnamefont {Uehlinger}}, \bibinfo {author}
  {\bibfnamefont {D.}~\bibnamefont {Greif}}, \ and\ \bibinfo {author}
  {\bibfnamefont {T.}~\bibnamefont {Esslinger}},\ }\href
  {http://dx.doi.org/10.1038/nature13915} {\bibfield  {journal} {\bibinfo
  {journal} {Nature}\ }\textbf {\bibinfo {volume} {515}},\ \bibinfo {pages}
  {237} (\bibinfo {year} {2014})}\BibitemShut {NoStop}%
\bibitem [{\citenamefont {Bilitewski}\ and\ \citenamefont
  {Cooper}(2015)}]{Bilitewski:pra2015}%
  \BibitemOpen
  \bibfield  {author} {\bibinfo {author} {\bibfnamefont {T.}~\bibnamefont
  {Bilitewski}}\ and\ \bibinfo {author} {\bibfnamefont {N.~R.}\ \bibnamefont
  {Cooper}},\ }\href {\doibase 10.1103/PhysRevA.91.063611} {\bibfield
  {journal} {\bibinfo  {journal} {Phys. Rev. A}\ }\textbf {\bibinfo {volume}
  {91}},\ \bibinfo {pages} {063611} (\bibinfo {year} {2015})}\BibitemShut
  {NoStop}%
\bibitem [{\citenamefont {Sentef}\ \emph {et~al.}(2015)\citenamefont {Sentef},
  \citenamefont {Claassen}, \citenamefont {Kemper}, \citenamefont {Moritz},
  \citenamefont {Oka}, \citenamefont {Freericks},\ and\ \citenamefont
  {Devereaux}}]{Sentef2014}%
  \BibitemOpen
  \bibfield  {author} {\bibinfo {author} {\bibfnamefont {M.~A.}\ \bibnamefont
  {Sentef}}, \bibinfo {author} {\bibfnamefont {M.}~\bibnamefont {Claassen}},
  \bibinfo {author} {\bibfnamefont {A.~F.}\ \bibnamefont {Kemper}}, \bibinfo
  {author} {\bibfnamefont {B.}~\bibnamefont {Moritz}}, \bibinfo {author}
  {\bibfnamefont {T.}~\bibnamefont {Oka}}, \bibinfo {author} {\bibfnamefont
  {J.~K.}\ \bibnamefont {Freericks}}, \ and\ \bibinfo {author} {\bibfnamefont
  {T.~P.}\ \bibnamefont {Devereaux}},\ }\href {\doibase 10.1038/ncomms8047}
  {\bibfield  {journal} {\bibinfo  {journal} {Nat. Commun.}\ }\textbf {\bibinfo
  {volume} {6}},\ \bibinfo {pages} {7047} (\bibinfo {year} {2015})}\BibitemShut
  {NoStop}%
\bibitem [{\citenamefont {Wang}\ \emph {et~al.}(2014)\citenamefont {Wang},
  \citenamefont {Steinberg}, \citenamefont {Jarillo-Herrero},\ and\
  \citenamefont {Gedik}}]{Wang2013}%
  \BibitemOpen
  \bibfield  {author} {\bibinfo {author} {\bibfnamefont {Y.}~\bibnamefont
  {Wang}}, \bibinfo {author} {\bibfnamefont {H.}~\bibnamefont {Steinberg}},
  \bibinfo {author} {\bibfnamefont {P.}~\bibnamefont {Jarillo-Herrero}}, \ and\
  \bibinfo {author} {\bibfnamefont {N.}~\bibnamefont {Gedik}},\ }\href
  {\doibase 10.1364/UP.2014.10.Thu.A.2} {\bibfield  {journal} {\bibinfo
  {journal} {19th Int. Conf. Ultrafast Phenom.}\ }\textbf {\bibinfo {volume}
  {108}},\ \bibinfo {pages} {10.Thu.A.2} (\bibinfo {year} {2014})},\ \Eprint
  {http://arxiv.org/abs/1310.7563} {1310.7563} \BibitemShut {NoStop}%
\bibitem [{\citenamefont {Mahmood}\ \emph {et~al.}(2016)\citenamefont
  {Mahmood}, \citenamefont {Chan}, \citenamefont {Alpichshev}, \citenamefont
  {Gardner}, \citenamefont {Lee}, \citenamefont {Lee},\ and\ \citenamefont
  {Gedik}}]{Mahmood:nat16}%
  \BibitemOpen
  \bibfield  {author} {\bibinfo {author} {\bibfnamefont {F.}~\bibnamefont
  {Mahmood}}, \bibinfo {author} {\bibfnamefont {C.-K.}\ \bibnamefont {Chan}},
  \bibinfo {author} {\bibfnamefont {Z.}~\bibnamefont {Alpichshev}}, \bibinfo
  {author} {\bibfnamefont {D.}~\bibnamefont {Gardner}}, \bibinfo {author}
  {\bibfnamefont {Y.}~\bibnamefont {Lee}}, \bibinfo {author} {\bibfnamefont
  {P.~A.}\ \bibnamefont {Lee}}, \ and\ \bibinfo {author} {\bibfnamefont
  {N.}~\bibnamefont {Gedik}},\ }\href {http://dx.doi.org/10.1038/nphys3609}
  {\bibfield  {journal} {\bibinfo  {journal} {Nat Phys}\ }\textbf {\bibinfo
  {volume} {12}},\ \bibinfo {pages} {306} (\bibinfo {year} {2016})}\BibitemShut
  {NoStop}%
\bibitem [{\citenamefont {Calvo}\ \emph {et~al.}(2015)\citenamefont {Calvo},
  \citenamefont {Foa~Torres}, \citenamefont {Perez-Piskunow}, \citenamefont
  {Balseiro},\ and\ \citenamefont {Usaj}}]{Calvo:prb15}%
  \BibitemOpen
  \bibfield  {author} {\bibinfo {author} {\bibfnamefont {H.~L.}\ \bibnamefont
  {Calvo}}, \bibinfo {author} {\bibfnamefont {L.~E.~F.}\ \bibnamefont
  {Foa~Torres}}, \bibinfo {author} {\bibfnamefont {P.~M.}\ \bibnamefont
  {Perez-Piskunow}}, \bibinfo {author} {\bibfnamefont {C.~A.}\ \bibnamefont
  {Balseiro}}, \ and\ \bibinfo {author} {\bibfnamefont {G.}~\bibnamefont
  {Usaj}},\ }\href {\doibase 10.1103/PhysRevB.91.241404} {\bibfield  {journal}
  {\bibinfo  {journal} {Phys. Rev. B}\ }\textbf {\bibinfo {volume} {91}},\
  \bibinfo {pages} {241404} (\bibinfo {year} {2015})}\BibitemShut {NoStop}%
\bibitem [{\citenamefont {Dal~Lago}\ \emph {et~al.}(2015)\citenamefont
  {Dal~Lago}, \citenamefont {Atala},\ and\ \citenamefont
  {Foa~Torres}}]{Dal:pra15}%
  \BibitemOpen
  \bibfield  {author} {\bibinfo {author} {\bibfnamefont {V.}~\bibnamefont
  {Dal~Lago}}, \bibinfo {author} {\bibfnamefont {M.}~\bibnamefont {Atala}}, \
  and\ \bibinfo {author} {\bibfnamefont {L.~E.~F.}\ \bibnamefont
  {Foa~Torres}},\ }\href {\doibase 10.1103/PhysRevA.92.023624} {\bibfield
  {journal} {\bibinfo  {journal} {Phys. Rev. A}\ }\textbf {\bibinfo {volume}
  {92}},\ \bibinfo {pages} {023624} (\bibinfo {year} {2015})}\BibitemShut
  {NoStop}%
\bibitem [{\citenamefont {Perez-Piskunow}\ \emph {et~al.}(2014)\citenamefont
  {Perez-Piskunow}, \citenamefont {Usaj}, \citenamefont {Balseiro},\ and\
  \citenamefont {Torres}}]{Perez:prb14}%
  \BibitemOpen
  \bibfield  {author} {\bibinfo {author} {\bibfnamefont {P.~M.}\ \bibnamefont
  {Perez-Piskunow}}, \bibinfo {author} {\bibfnamefont {G.}~\bibnamefont
  {Usaj}}, \bibinfo {author} {\bibfnamefont {C.~A.}\ \bibnamefont {Balseiro}},
  \ and\ \bibinfo {author} {\bibfnamefont {L.~E. F.~F.}\ \bibnamefont
  {Torres}},\ }\href {\doibase 10.1103/PhysRevB.89.121401} {\bibfield
  {journal} {\bibinfo  {journal} {Phys. Rev. B}\ }\textbf {\bibinfo {volume}
  {89}},\ \bibinfo {pages} {121401} (\bibinfo {year} {2014})}\BibitemShut
  {NoStop}%
\bibitem [{\citenamefont {Perez-Piskunow}\ \emph {et~al.}(2015)\citenamefont
  {Perez-Piskunow}, \citenamefont {Foa~Torres},\ and\ \citenamefont
  {Usaj}}]{Perez:pra15}%
  \BibitemOpen
  \bibfield  {author} {\bibinfo {author} {\bibfnamefont {P.~M.}\ \bibnamefont
  {Perez-Piskunow}}, \bibinfo {author} {\bibfnamefont {L.~E.~F.}\ \bibnamefont
  {Foa~Torres}}, \ and\ \bibinfo {author} {\bibfnamefont {G.}~\bibnamefont
  {Usaj}},\ }\href {\doibase 10.1103/PhysRevA.91.043625} {\bibfield  {journal}
  {\bibinfo  {journal} {Phys. Rev. A}\ }\textbf {\bibinfo {volume} {91}},\
  \bibinfo {pages} {043625} (\bibinfo {year} {2015})}\BibitemShut {NoStop}%
\bibitem [{\citenamefont {Kim}\ \emph {et~al.}(2014)\citenamefont {Kim},
  \citenamefont {Ikeda},\ and\ \citenamefont {Huse}}]{Hyungwon:pre14}%
  \BibitemOpen
  \bibfield  {author} {\bibinfo {author} {\bibfnamefont {H.}~\bibnamefont
  {Kim}}, \bibinfo {author} {\bibfnamefont {T.~N.}\ \bibnamefont {Ikeda}}, \
  and\ \bibinfo {author} {\bibfnamefont {D.~A.}\ \bibnamefont {Huse}},\ }\href
  {\doibase 10.1103/PhysRevE.90.052105} {\bibfield  {journal} {\bibinfo
  {journal} {Phys. Rev. E}\ }\textbf {\bibinfo {volume} {90}},\ \bibinfo
  {pages} {052105} (\bibinfo {year} {2014})}\BibitemShut {NoStop}%
\bibitem [{\citenamefont {Ponte}\ \emph {et~al.}(2015)\citenamefont {Ponte},
  \citenamefont {Papi\ifmmode~\acute{c}\else \'{c}\fi{}}, \citenamefont
  {Huveneers},\ and\ \citenamefont {Abanin}}]{Ponte:prl15}%
  \BibitemOpen
  \bibfield  {author} {\bibinfo {author} {\bibfnamefont {P.}~\bibnamefont
  {Ponte}}, \bibinfo {author} {\bibfnamefont {Z.}~\bibnamefont
  {Papi\ifmmode~\acute{c}\else \'{c}\fi{}}}, \bibinfo {author} {\bibfnamefont
  {F.~m.~c.}\ \bibnamefont {Huveneers}}, \ and\ \bibinfo {author}
  {\bibfnamefont {D.~A.}\ \bibnamefont {Abanin}},\ }\href {\doibase
  10.1103/PhysRevLett.114.140401} {\bibfield  {journal} {\bibinfo  {journal}
  {Phys. Rev. Lett.}\ }\textbf {\bibinfo {volume} {114}},\ \bibinfo {pages}
  {140401} (\bibinfo {year} {2015})}\BibitemShut {NoStop}%
\bibitem [{\citenamefont {Lazarides}\ \emph {et~al.}(2015)\citenamefont
  {Lazarides}, \citenamefont {Das},\ and\ \citenamefont
  {Moessner}}]{Lazarides:prl15}%
  \BibitemOpen
  \bibfield  {author} {\bibinfo {author} {\bibfnamefont {A.}~\bibnamefont
  {Lazarides}}, \bibinfo {author} {\bibfnamefont {A.}~\bibnamefont {Das}}, \
  and\ \bibinfo {author} {\bibfnamefont {R.}~\bibnamefont {Moessner}},\ }\href
  {\doibase 10.1103/PhysRevLett.115.030402} {\bibfield  {journal} {\bibinfo
  {journal} {Phys. Rev. Lett.}\ }\textbf {\bibinfo {volume} {115}},\ \bibinfo
  {pages} {030402} (\bibinfo {year} {2015})}\BibitemShut {NoStop}%
\bibitem [{\citenamefont {Genske}\ and\ \citenamefont
  {Rosch}(2015)}]{Genske:pra15}%
  \BibitemOpen
  \bibfield  {author} {\bibinfo {author} {\bibfnamefont {M.}~\bibnamefont
  {Genske}}\ and\ \bibinfo {author} {\bibfnamefont {A.}~\bibnamefont {Rosch}},\
  }\href {\doibase 10.1103/PhysRevA.92.062108} {\bibfield  {journal} {\bibinfo
  {journal} {Phys. Rev. A}\ }\textbf {\bibinfo {volume} {92}},\ \bibinfo
  {pages} {062108} (\bibinfo {year} {2015})}\BibitemShut {NoStop}%
\bibitem [{\citenamefont {Dehghani}\ \emph {et~al.}(2014)\citenamefont
  {Dehghani}, \citenamefont {Oka},\ and\ \citenamefont {Mitra}}]{Dehghani2014}%
  \BibitemOpen
  \bibfield  {author} {\bibinfo {author} {\bibfnamefont {H.}~\bibnamefont
  {Dehghani}}, \bibinfo {author} {\bibfnamefont {T.}~\bibnamefont {Oka}}, \
  and\ \bibinfo {author} {\bibfnamefont {A.}~\bibnamefont {Mitra}},\ }\href
  {\doibase 10.1103/PhysRevB.90.195429} {\bibfield  {journal} {\bibinfo
  {journal} {Phys. Rev. B}\ }\textbf {\bibinfo {volume} {90}},\ \bibinfo
  {pages} {195429} (\bibinfo {year} {2014})}\BibitemShut {NoStop}%
\bibitem [{\citenamefont {Dehghani}\ and\ \citenamefont
  {Mitra}(2016)}]{Dehghani:prb16}%
  \BibitemOpen
  \bibfield  {author} {\bibinfo {author} {\bibfnamefont {H.}~\bibnamefont
  {Dehghani}}\ and\ \bibinfo {author} {\bibfnamefont {A.}~\bibnamefont
  {Mitra}},\ }\href {\doibase 10.1103/PhysRevB.93.245416} {\bibfield  {journal}
  {\bibinfo  {journal} {Phys. Rev. B}\ }\textbf {\bibinfo {volume} {93}},\
  \bibinfo {pages} {245416} (\bibinfo {year} {2016})}\BibitemShut {NoStop}%
\bibitem [{\citenamefont {Iadecola}\ and\ \citenamefont
  {Chamon}(2015)}]{Iadecola:prb15}%
  \BibitemOpen
  \bibfield  {author} {\bibinfo {author} {\bibfnamefont {T.}~\bibnamefont
  {Iadecola}}\ and\ \bibinfo {author} {\bibfnamefont {C.}~\bibnamefont
  {Chamon}},\ }\href {\doibase 10.1103/PhysRevB.91.184301} {\bibfield
  {journal} {\bibinfo  {journal} {Phys. Rev. B}\ }\textbf {\bibinfo {volume}
  {91}},\ \bibinfo {pages} {184301} (\bibinfo {year} {2015})}\BibitemShut
  {NoStop}%
\bibitem [{\citenamefont {Iadecola}\ \emph {et~al.}(2015)\citenamefont
  {Iadecola}, \citenamefont {Neupert},\ and\ \citenamefont
  {Chamon}}]{Iadecola:prb15a}%
  \BibitemOpen
  \bibfield  {author} {\bibinfo {author} {\bibfnamefont {T.}~\bibnamefont
  {Iadecola}}, \bibinfo {author} {\bibfnamefont {T.}~\bibnamefont {Neupert}}, \
  and\ \bibinfo {author} {\bibfnamefont {C.}~\bibnamefont {Chamon}},\ }\href
  {\doibase 10.1103/PhysRevB.91.235133} {\bibfield  {journal} {\bibinfo
  {journal} {Phys. Rev. B}\ }\textbf {\bibinfo {volume} {91}},\ \bibinfo
  {pages} {235133} (\bibinfo {year} {2015})}\BibitemShut {NoStop}%
\bibitem [{\citenamefont {Seetharam}\ \emph {et~al.}(2015)\citenamefont
  {Seetharam}, \citenamefont {Bardyn}, \citenamefont {Lindner}, \citenamefont
  {Rudner},\ and\ \citenamefont {Refael}}]{Seetharam:prx15}%
  \BibitemOpen
  \bibfield  {author} {\bibinfo {author} {\bibfnamefont {K.~I.}\ \bibnamefont
  {Seetharam}}, \bibinfo {author} {\bibfnamefont {C.-E.}\ \bibnamefont
  {Bardyn}}, \bibinfo {author} {\bibfnamefont {N.~H.}\ \bibnamefont {Lindner}},
  \bibinfo {author} {\bibfnamefont {M.~S.}\ \bibnamefont {Rudner}}, \ and\
  \bibinfo {author} {\bibfnamefont {G.}~\bibnamefont {Refael}},\ }\href
  {\doibase 10.1103/PhysRevX.5.041050} {\bibfield  {journal} {\bibinfo
  {journal} {Phys. Rev. X}\ }\textbf {\bibinfo {volume} {5}},\ \bibinfo {pages}
  {041050} (\bibinfo {year} {2015})}\BibitemShut {NoStop}%
\bibitem [{\citenamefont {Shirai}\ \emph {et~al.}(2015)\citenamefont {Shirai},
  \citenamefont {Mori},\ and\ \citenamefont {Miyashita}}]{Shirai:pre15}%
  \BibitemOpen
  \bibfield  {author} {\bibinfo {author} {\bibfnamefont {T.}~\bibnamefont
  {Shirai}}, \bibinfo {author} {\bibfnamefont {T.}~\bibnamefont {Mori}}, \ and\
  \bibinfo {author} {\bibfnamefont {S.}~\bibnamefont {Miyashita}},\ }\href
  {\doibase 10.1103/PhysRevE.91.030101} {\bibfield  {journal} {\bibinfo
  {journal} {Phys. Rev. E}\ }\textbf {\bibinfo {volume} {91}},\ \bibinfo
  {pages} {030101} (\bibinfo {year} {2015})}\BibitemShut {NoStop}%
\bibitem [{\citenamefont {Tsuji}\ \emph {et~al.}(2009)\citenamefont {Tsuji},
  \citenamefont {Oka},\ and\ \citenamefont {Aoki}}]{Tsuji2009}%
  \BibitemOpen
  \bibfield  {author} {\bibinfo {author} {\bibfnamefont {N.}~\bibnamefont
  {Tsuji}}, \bibinfo {author} {\bibfnamefont {T.}~\bibnamefont {Oka}}, \ and\
  \bibinfo {author} {\bibfnamefont {H.}~\bibnamefont {Aoki}},\ }\href {\doibase
  10.1103/PhysRevLett.103.047403} {\bibfield  {journal} {\bibinfo  {journal}
  {Phys. Rev. Lett.}\ }\textbf {\bibinfo {volume} {103}},\ \bibinfo {pages} {1}
  (\bibinfo {year} {2009})}\BibitemShut {NoStop}%
\bibitem [{\citenamefont {Dehghani}\ \emph {et~al.}(2015)\citenamefont
  {Dehghani}, \citenamefont {Oka},\ and\ \citenamefont {Mitra}}]{Dehghani2015}%
  \BibitemOpen
  \bibfield  {author} {\bibinfo {author} {\bibfnamefont {H.}~\bibnamefont
  {Dehghani}}, \bibinfo {author} {\bibfnamefont {T.}~\bibnamefont {Oka}}, \
  and\ \bibinfo {author} {\bibfnamefont {A.}~\bibnamefont {Mitra}},\ }\href
  {\doibase 10.1103/PhysRevB.91.155422} {\bibfield  {journal} {\bibinfo
  {journal} {Phys. Rev. B}\ }\textbf {\bibinfo {volume} {91}},\ \bibinfo
  {pages} {1} (\bibinfo {year} {2015})}\BibitemShut {NoStop}%
\bibitem [{\citenamefont {Dehghani}\ and\ \citenamefont
  {Mitra}(2015)}]{Dehghani2015a}%
  \BibitemOpen
  \bibfield  {author} {\bibinfo {author} {\bibfnamefont {H.}~\bibnamefont
  {Dehghani}}\ and\ \bibinfo {author} {\bibfnamefont {A.}~\bibnamefont
  {Mitra}},\ }\href {\doibase 10.1103/PhysRevB.92.165111} {\bibfield  {journal}
  {\bibinfo  {journal} {Phys. Rev. B}\ }\textbf {\bibinfo {volume} {92}},\
  \bibinfo {pages} {1} (\bibinfo {year} {2015})}\BibitemShut {NoStop}%
\bibitem [{\citenamefont {Floquet}(1883)}]{GFloquet1883}%
  \BibitemOpen
  \bibfield  {author} {\bibinfo {author} {\bibfnamefont {G.}~\bibnamefont
  {Floquet}},\ }\href@noop {} {\bibfield  {journal} {\bibinfo  {journal} {Ann.
  Sci. Ec. Normale Super.}\ }\textbf {\bibinfo {volume} {12}},\ \bibinfo
  {pages} {47} (\bibinfo {year} {1883})}\BibitemShut {NoStop}%
\bibitem [{\citenamefont {Bukov}\ \emph {et~al.}(2015)\citenamefont {Bukov},
  \citenamefont {D'Alessio},\ and\ \citenamefont {Polkovnikov}}]{Bukov}%
  \BibitemOpen
  \bibfield  {author} {\bibinfo {author} {\bibfnamefont {M.}~\bibnamefont
  {Bukov}}, \bibinfo {author} {\bibfnamefont {L.}~\bibnamefont {D'Alessio}}, \
  and\ \bibinfo {author} {\bibfnamefont {A.}~\bibnamefont {Polkovnikov}},\
  }\href {\doibase 10.1080/00018732.2015.1055918} {\bibfield  {journal}
  {\bibinfo  {journal} {Adv. Phys.}\ }\textbf {\bibinfo {volume} {64}},\
  \bibinfo {pages} {139} (\bibinfo {year} {2015})}\BibitemShut {NoStop}%
\bibitem [{\citenamefont {Freericks}\ \emph {et~al.}(2009)\citenamefont
  {Freericks}, \citenamefont {Krishnamurthy},\ and\ \citenamefont
  {Pruschke}}]{Freericks2009a}%
  \BibitemOpen
  \bibfield  {author} {\bibinfo {author} {\bibfnamefont {J.~K.}\ \bibnamefont
  {Freericks}}, \bibinfo {author} {\bibfnamefont {H.~R.}\ \bibnamefont
  {Krishnamurthy}}, \ and\ \bibinfo {author} {\bibfnamefont {T.}~\bibnamefont
  {Pruschke}},\ }\href {\doibase 10.1103/PhysRevLett.102.136401} {\bibfield
  {journal} {\bibinfo  {journal} {Phys. Rev. Lett.}\ }\textbf {\bibinfo
  {volume} {102}},\ \bibinfo {pages} {3} (\bibinfo {year} {2009})}\BibitemShut
  {NoStop}%
\end{thebibliography}%


\appendix

\section{DERIVATION OF LONGITUDINAL OPTICAL CONDUCTIVITY}
\label{sec:DERIVATION_OF_LONGITUDINAL_OPTICAL_CONDUCTIVITY}

In this section we derive the general form of the optical conductivity\cite{Dehghani2015a}. The current-current correlation function which quantifies how an electric field applied in the direction \(\hat{i}\) affects the current flowing
in the direction \(\hat{i}\) is given by
\begin{equation}
R_{i i}^C\left(\mathbf{q},t,t'\right)=-i\left\langle T_C\left[J_{\mathbf{q} I}^i(t)J_{-\mathbf{q} I}^i\left(t'\right)\right]\right\rangle, 
\label{eq: Response}
\end{equation}
where
\begin{equation}
J_{\mathbf{q}}^i(t)=\frac{1}{\sqrt{N}}\sum _{\mathbf{k},\sigma \sigma '} c_{\mathbf{k}+\mathbf{q}/2, \sigma }^{\dagger }(t)c_{\mathbf{k}-\mathbf{q}/2, \sigma '}(t)\frac{\partial
h_{\mathbf{k}}^{\sigma \sigma '}(t)}{\partial k_i},
\end{equation}
is the current operator in the interaction representation evolved from $t=t_0$: 
\begin{equation}
\mathbf{J}_{\mathbf{k} I}(t)=U_{\mathbf{k}}\left(t_0,t\right)\mathbf{J}_{\mathbf{k}}\left(t_0\right)U_{\mathbf{k}}\left(t,t_0\right).
\end{equation}
The time-evolution operator is given by,
\begin{equation}
U_{\mathbf{k}}\left(t,t_0\right)=\sum _{\alpha } e^{-i \epsilon _{\mathbf{k}\alpha }\left(t-t_0\right)}|\phi _{\mathbf{k}\alpha }(t)\rangle \langle \phi _{\mathbf{k}\alpha
}\left(t_0\right)|, 
\label{eq: evol_operator}
\end{equation}
where \(\epsilon _{\mathbf{k}\alpha }\) is the quasi-energy and
\begin{equation}
|\phi _{\mathbf{k}\alpha }(t)\rangle =\left(
\begin{array}{c}
 \phi _{\mathbf{k}\alpha }^{\text{up}}(t) \\
 \phi _{\mathbf{k}\alpha }^{\text{dn}}(t) \\
\end{array}
\right),
\end{equation}
is the Floquet eigenvector. Thus, Eq. (\ref{eq: evol_operator}) becomes
\begin{equation}
U_{\mathbf{k}\sigma \sigma '}\left(t,t_0\right)=\sum _{\alpha } e^{-i \epsilon _{\mathbf{k}\alpha }\left(t-t_0\right)}\phi _{\mathbf{k}\alpha }^{\sigma }(t)\phi
_{\mathbf{k}\alpha }^{\sigma '*}\left(t_0\right). 
\end{equation}
In the interaction representation, 
\begin{equation}
c_{\mathbf{k}\sigma }^I(t)=U_{\mathbf{k}\sigma \sigma '}\left(t,t_0\right)c_{\mathbf{k}\sigma '}^I\left(t_0\right),
\label{eq: fermi_annihilate}
\end{equation}
\begin{equation}
c_{\mathbf{k}\sigma }^{I \dagger }(t)=c_{\mathbf{k}\sigma '}^{I \dagger }\left(t_0\right)U_{\mathbf{k}\sigma '\sigma }\left(t_0,t\right).
\label{eq: fermi_create}
\end{equation}
We expand the fermionic operators in the quasi-mode basis at time \(t_0\) as
\begin{equation}
c_{\mathbf{k}\sigma }^I\left(t_0\right)=\sum _{\alpha '} \phi _{\mathbf{k}\alpha '}^{\sigma }\left(t_0\right)\gamma _{\mathbf{k}\alpha '}.
\label{eq: c_to_gamma}
\end{equation}
By combining Eqs.(\ref{eq: fermi_annihilate}-\ref{eq: c_to_gamma}) and then inserting the result into Eq.(\ref{eq: Response}), the response function becomes
\begin{widetext}
\begin{eqnarray}
R_{i j}\left(\mathbf{q},t,t'\right) & = & -i \theta (t-t')\frac{1}{N}\sum _{\mathbf{k},\alpha \beta \gamma \delta } e^{-i\left(\epsilon _{\mathbf{k}\mathbf{-}\frac{\mathbf{q}}{2}\alpha
}-\epsilon _{\mathbf{k}\mathbf{+}\frac{\mathbf{q}}{2}\beta }\right)\left(\bar{t}+\frac{t_r}{2}-t_0\right)}e^{-i\left(\epsilon _{\mathbf{k}\mathbf{+}\frac{\mathbf{q}}{2}\gamma
}-\epsilon _{\mathbf{k}\mathbf{-}\frac{\mathbf{q}}{2}\delta }\right)\left(\bar{t}-\frac{t_r}{2}-t_0\right)} \nonumber \\
& & \times \langle \phi _{\mathbf{k}\mathbf{+}\frac{\mathbf{q}}{2}\beta }(t)|\left[\frac{\partial h_{\mathbf{k}}(t)}{\partial k_i}\right]|\phi _{\mathbf{k}\mathbf{-}\frac{\mathbf{q}}{2}\alpha
}(t)\rangle \langle \phi _{\mathbf{k}\mathbf{-}\frac{\mathbf{q}}{2}\delta }\left(t'\right)|\left[\frac{\partial h_{\mathbf{k}}\left(t'\right)}{\partial k_j}\right]|\phi
_{\mathbf{k}\mathbf{+}\frac{\mathbf{q}}{2}\gamma }\left(t'\right)\rangle \nonumber \\
& & \times \langle \Psi \left(t_0\right)|\left[\gamma _{\mathbf{k}\mathbf{+}\frac{\mathbf{q}}{2}\beta
}^{\dagger }\gamma _{\mathbf{k}\mathbf{-}\frac{\mathbf{q}}{2}\alpha },\gamma _{\mathbf{k}\mathbf{-}\frac{\mathbf{q}}{2}\delta }^{\dagger }\gamma _{\mathbf{k}\mathbf{+}\frac{\mathbf{q}}{2}\gamma
}]\right.|\Psi \left(t_0\right)\rangle \nonumber \\
& \approx & -i \theta (t-t')\frac{1}{N}\sum _{\mathbf{k},\alpha \beta } e^{-i\left(\epsilon _{\mathbf{k}\mathbf{-}\frac{\mathbf{q}}{2}\alpha }-\epsilon _{\mathbf{k}\mathbf{+}\frac{\mathbf{q}}{2}\beta
}\right)t_r} \nonumber \\
& & \times \langle \phi _{\mathbf{k}\mathbf{+}\frac{\mathbf{q}}{2}\beta }(t)|\left[\frac{\partial h_{\mathbf{k}}(t)}{\partial k_i}\right]|\phi _{\mathbf{k}\mathbf{-}\frac{\mathbf{q}}{2}\alpha
}(t)\rangle \langle \phi _{\mathbf{k}\mathbf{-}\frac{\mathbf{q}}{2}\alpha }\left(t'\right)|\left[\frac{\partial h_{\mathbf{k}}\left(t'\right)}{\partial k_j}\right]|\phi
_{\mathbf{k}\mathbf{+}\frac{\mathbf{q}}{2}\beta }\left(t'\right)\rangle \nonumber \\
& & \times \langle \Psi \left(t_0\right)|\gamma _{\mathbf{k}\mathbf{+}\frac{\mathbf{q}}{2}\beta
}^{\dagger }\gamma _{\mathbf{k}\mathbf{+}\frac{\mathbf{q}}{2}\beta }-\gamma _{\mathbf{k}\mathbf{-}\frac{\mathbf{q}}{2}\alpha }^{\dagger }\gamma _{\mathbf{k}\mathbf{-}\frac{\mathbf{q}}{2}\alpha
}|\Psi \left(t_0\right)\rangle,
\label{eq: Response_approx}
\end{eqnarray}
\end{widetext}
where \(\alpha ,\beta =u,d\), \(u,d\) represent upper and lower band in a Floquet mode respectively. In the last approximate equality of Eq.(\ref{eq: Response_approx}), we drop the term with fast oscillation factor $e^{-i (\epsilon_{k u}-\epsilon_{k d})(t+t')/2}$. We set
\begin{equation}
\langle \phi _{\mathbf{k}\mathbf{+}\frac{\mathbf{q}}{2}\beta }(t)|\left[\frac{\partial h_{\mathbf{k}}(t)}{\partial k_i}\right]|\phi _{\mathbf{k}\mathbf{-}\frac{\mathbf{q}}{2}\alpha
}(t)\rangle =\sum _m e^{i m \Omega  t}D_{\beta  i \alpha }^m(\mathbf{k,q}),
\end{equation}
and rewrite Eq.(\ref{eq: Response_approx}) as
\begin{eqnarray}
R_{i j}\left(\mathbf{q},t,t'\right) &=& -i \theta (t-t')\frac{1}{N}\sum _{\mathbf{k},\alpha \beta } e^{-i\left(\epsilon _{\mathbf{k}\mathbf{-}\frac{\mathbf{q}}{2}\alpha
}-\epsilon _{\mathbf{k}\mathbf{+}\frac{\mathbf{q}}{2}\beta }\right)t_r} \nonumber \\
&&\times \sum _m e^{i m \Omega  t}D_{\beta  i \alpha }^m(\mathbf{k,q})\sum _{m'} e^{i m' \Omega  t'}D_{\alpha  j \beta }^{m'}(\mathbf{k}\mathbf{,}-\mathbf{q}) \nonumber \\
&&\times
\langle \Psi \left(t_0\right)|\gamma _{\mathbf{k}\mathbf{+}\frac{\mathbf{q}}{2}\beta }^{\dagger }\gamma _{\mathbf{k}\mathbf{+}\frac{\mathbf{q}}{2}\beta }-\gamma _{\mathbf{k}\mathbf{-}\frac{\mathbf{q}}{2}\alpha
}^{\dagger }\gamma _{\mathbf{k}\mathbf{-}\frac{\mathbf{q}}{2}\alpha }|\Psi \left(t_0\right)\rangle. \nonumber \\
\label{eq: Response_approx_tr}
\end{eqnarray}
Averaged over $\frac{t+t'}{2}$, only \(m=-m'\) term of Eq.(\ref{eq: Response_approx_tr}) has a contribution:
\begin{widetext}
\begin{eqnarray}
R_{i j}\left(\mathbf{q},t_r,\text{mode}=0\right)&=&-i \theta (t_r)\frac{1}{N}\sum _{\mathbf{k},\alpha \beta } e^{-i\left(\epsilon _{\mathbf{k}\mathbf{-}\frac{\mathbf{q}}{2}\alpha
}-\epsilon _{\mathbf{k}\mathbf{+}\frac{\mathbf{q}}{2}\beta }\right)t_r} \nonumber \\
&&\times \sum _m e^{i m \Omega  t_r}D_{\beta  i \alpha }^m(\mathbf{k,q})D_{\alpha  j \beta }^{-m}(\mathbf{k}\mathbf{,}-\mathbf{q})\langle \Psi \left(t_0\right)|[\gamma
_{\mathbf{k}\mathbf{+}\frac{\mathbf{q}}{2}\beta }^{\dagger }\gamma _{\mathbf{k}\mathbf{+}\frac{\mathbf{q}}{2}\beta }-\gamma _{\mathbf{k}\mathbf{-}\frac{\mathbf{q}}{2}\alpha
}^{\dagger }\gamma _{\mathbf{k}\mathbf{-}\frac{\mathbf{q}}{2}\alpha }]|\Psi \left(t_0\right)\rangle.
\label{eq: Response_approx_tr_0}
\end{eqnarray}
By Fourier transform Eq.(\ref{eq: Response_approx_tr_0}) with respect to \(t_r\), one arrives at
\begin{eqnarray}
R_{i j}(\mathbf{q},\omega ,\text{mode}=0)&=&\int dt_rR_{i j}\left(\mathbf{q},t_r,\text{mode}=0\right)e^{i (\omega +\text{i$\delta $}) t_r} \nonumber \\
&=&-i \int dt_re^{i (\omega +\text{i$\delta $}) t_r}\theta (t_r)\frac{1}{N}\sum _{\mathbf{k},\alpha \beta } e^{-i\left(\epsilon _{\mathbf{k}\mathbf{-}\frac{\mathbf{q}}{2}\alpha
}-\epsilon _{\mathbf{k}\mathbf{+}\frac{\mathbf{q}}{2}\beta }\right)t_r}\sum _m e^{i m \Omega  t_r}D_{\beta  i \alpha }^m(\mathbf{k,q})D_{\alpha  j \beta }^{-m}(\mathbf{k}\mathbf{,}-\mathbf{q}) \nonumber \\
&& \times \langle
\Psi \left(t_0\right)|[\gamma _{\mathbf{k}\mathbf{+}\frac{\mathbf{q}}{2}\beta }^{\dagger }\gamma _{\mathbf{k}\mathbf{+}\frac{\mathbf{q}}{2}\beta }-\gamma _{\mathbf{k}\mathbf{-}\frac{\mathbf{q}}{2}\alpha
}^{\dagger }\gamma _{\mathbf{k}\mathbf{-}\frac{\mathbf{q}}{2}\alpha }]|\Psi \left(t_0\right)\rangle \nonumber \\
&=&\frac{1}{N}\sum _{\mathbf{k},\alpha \beta } \sum _m \frac{D_{\beta  i \alpha }^m(\mathbf{k,q})D_{\alpha  j \beta }^{-m}(\mathbf{k}\mathbf{,}-\mathbf{q})\langle
\Psi \left(t_0\right)|[\gamma _{\mathbf{k}\mathbf{+}\frac{\mathbf{q}}{2}\beta }^{\dagger }\gamma _{\mathbf{k}\mathbf{+}\frac{\mathbf{q}}{2}\beta }-\gamma _{\mathbf{k}\mathbf{-}\frac{\mathbf{q}}{2}\alpha
}^{\dagger }\gamma _{\mathbf{k}\mathbf{-}\frac{\mathbf{q}}{2}\alpha }]|\Psi \left(t_0\right)\rangle }{\omega +\text{i$\delta $}-\left(\epsilon _{\mathbf{k}\mathbf{-}\frac{\mathbf{q}}{2}\alpha
}-\epsilon _{\mathbf{k}\mathbf{+}\frac{\mathbf{q}}{2}\beta }-m \Omega \right)}, 
\end{eqnarray}
where the longitudinal component can be extracted as
\begin{equation}
R_{i i}(\mathbf{q},\omega ,\text{mode}=0)=\frac{1}{N}\sum _{\mathbf{k},\alpha \beta } \sum _m \frac{D_{\beta  i \alpha }^m(\mathbf{k,q})D_{\alpha  i \beta
}^{-m}(\mathbf{k}\mathbf{,}-\mathbf{q})\langle \Psi \left(t_0\right)|[\gamma _{\mathbf{k}\mathbf{+}\frac{\mathbf{q}}{2}\beta }^{\dagger }\gamma _{\mathbf{k}\mathbf{+}\frac{\mathbf{q}}{2}\beta
}-\gamma _{\mathbf{k}\mathbf{-}\frac{\mathbf{q}}{2}\alpha }^{\dagger }\gamma _{\mathbf{k}\mathbf{-}\frac{\mathbf{q}}{2}\alpha }]|\Psi \left(t_0\right)\rangle }{\omega
+\text{i$\delta $}-\left(\epsilon _{\mathbf{k}\mathbf{-}\frac{\mathbf{q}}{2}\alpha }-\epsilon _{\mathbf{k}\mathbf{+}\frac{\mathbf{q}}{2}\beta }-m \Omega \right)}.
\label{eq: Response_long}
\end{equation}
In the limit \(\mathbf{q}\to 0\), Eq.(\ref{eq: Response_long}) is reduced to
\begin{eqnarray}
R_{i i}(\mathbf{q}=0,\omega ,\text{mode}=0)&=&\frac{1}{N}\sum _{\mathbf{k},\alpha \beta } \sum _m \frac{D_{\beta  i \alpha }^m(\mathbf{k})D_{\alpha  i \beta
}^{-m}(\mathbf{k})\langle \Psi \left(t_0\right)|[\gamma _{\mathbf{k}\mathbf{+}\frac{\mathbf{q}}{2}\beta }^{\dagger }\gamma _{\mathbf{k}\mathbf{+}\frac{\mathbf{q}}{2}\beta
}-\gamma _{\mathbf{k}\mathbf{-}\frac{\mathbf{q}}{2}\alpha }^{\dagger }\gamma _{\mathbf{k}\mathbf{-}\frac{\mathbf{q}}{2}\alpha }]|\Psi \left(t_0\right)\rangle }{\omega
+\text{i$\delta $}-\left(\epsilon _{\mathbf{k}\mathbf{-}\frac{\mathbf{q}}{2}\alpha }-\epsilon _{\mathbf{k}\mathbf{+}\frac{\mathbf{q}}{2}\beta }-m \Omega \right)} \nonumber \\
&=&\frac{1}{N}\sum _{\mathbf{k}} \sum _m D_{u i d}^mD_{d i u}^{-m}\langle \Psi \left(t_0\right)|[\gamma _{\mathbf{k}u}^{\dagger }\gamma _{\mathbf{k}u}-\gamma
_{\mathbf{k}d}^{\dagger }\gamma _{\mathbf{k}d}]|\Psi \left(t_0\right)\rangle \frac{2\left(\epsilon _{\mathbf{k}d}-\epsilon _{\mathbf{k}u}-m \Omega \right)}{\omega
^2-\left(\epsilon _{\mathbf{k}d}-\epsilon _{\mathbf{k}u}-m \Omega \right){}^2+2i \omega  \delta -\delta ^2} \nonumber \\
\end{eqnarray}
with
\begin{equation}
D_{u i d}^m(\mathbf{k})=\sum _{n l} \langle \tilde{\phi }_{\mathbf{k}u}^n|[\frac{\partial h_{\mathbf{k}}^{m+n-l}}{\partial k_i}]|\tilde{\phi }_{\mathbf{k}d}^l\rangle.
\end{equation}
Thus the longitudinal optical conductivity is evaluated as
\begin{eqnarray}
\text{Re} [\sigma _{i i}(\omega )]&\equiv& \frac{\text{Im} R_{i i}(\mathbf{q}=0,\omega ,\text{mode}=0)}{\omega } \nonumber \\
&=&\frac{1}{N}\sum _{\mathbf{k}} \sum _m D_{u i d}^m(\mathbf{k})D_{d i u}^{-m}(\mathbf{k})(\rho _{\mathbf{k}u}-\rho _{\mathbf{k}d}) \nonumber \\
&& \times \frac{-4\left(\epsilon _{\mathbf{k}d}-\epsilon _{\mathbf{k}u}-m
\Omega \right)\delta }{\left[\omega ^2-\left(\epsilon _{\mathbf{k}d}-\epsilon _{\mathbf{k}u}-m \Omega \right){}^2\right]{}^2+2\left(\omega ^2+\left(\epsilon
_{\mathbf{k}d}-\epsilon _{\mathbf{k}u}-m \Omega \right){}^2\right) \delta ^2}.
\end{eqnarray}
\end{widetext}

\section{LOW ENERGY EFFECTIVE HAMILTONIAN}
\label{sec:LOW_ENERGY_EFFECTIVE_HAMILTONIAN}

In this section, we derive the low energy time-dependent Hamiltonian from the lattice model Eq.\eqref{eq: ham_stat} which we rewrite here for convenience:
\begin{widetext}
\begin{eqnarray}
h_{\mathbf{k}}^{A B}\left(t_1\right)&=&2t e^{i \mathbf{A}\left(t_1\right)\cdot \mathbf{\delta }_3}+\sum _{i=1,2} t e^{i \mathbf{k}\mathbf{\cdot }\mathbf{a}_i+i \mathbf{A}\left(t_1\right)\cdot
\mathbf{\delta }_i}\nonumber \\
&=& t e^{i \left[\frac{3k_x}{2}+\frac{\sqrt{3}k_y}{2}+\frac{A_x\left(t_1\right)}{2}+\frac{\sqrt{3}A_y\left(t_1\right)}{2}\right]}+t e^{i \left[\frac{3k_x}{2}-\frac{\sqrt{3}k_y}{2}+\frac{A_x\left(t_1\right)}{2}-\frac{\sqrt{3}A_y\left(t_1\right)}{2}\right]}+2t
e^{-i A_x\left(t_1\right)},
\end{eqnarray}
\end{widetext}
where we used \(t_1\) as time to be distinguished from the hopping parameter. By expanding Eq.(\ref{eq: ham_n}) up to \(O\left(k^2\right)\) and \(O\left(A^2\right)\)
in the vicinity of \(\mathbf{D}=\left(\frac{2\pi }{3},0\right)\), one arrives at
\begin{widetext}
\begin{equation}
h_{\mathbf{k}}^{A B}\left(t_1\right)\approx -3t \left(A_x\left(t_1\right)+p_x\right)i-\frac{3t}{4} \left(A_x\left(t_1\right){}^2-A_y\left(t_1\right){}^2-2
A_x\left(t_1\right) p_x-3 p_x{}^2-2 A_y\left(t_1\right) p_y-p_y{}^2\right),
\end{equation}
where \(\left(p_x,p_y\right)\) is the momentum around \(\mathbf{D}=\left(\frac{2\pi }{3},0\right)\).  This can also be written in the compact matrix form
as
\begin{equation}
H_{\mathbf{k}}\left(t_1\right)=\frac{3t}{4} \left(-A_x\left(t_1\right){}^2+A_y\left(t_1\right){}^2+2 A_x\left(t_1\right) p_x+3 p_x{}^2+2 A_y\left(t_1\right)
p_y+p_y{}^2\right)\sigma _x+3t \left(A_x\left(t_1\right)+p_x\right)\sigma _y.
\end{equation}
\end{widetext}
The dominant features of the band structure can be understood by considering the effective Hamiltonian at large driving frequency $\Omega $, which is given by\cite{Bukov}
\begin{widetext}
\begin{eqnarray}
H_{\mathbf{k}}^{\text{eff}}&=&H_{\mathbf{k}}^0+\frac{1}{\Omega }\left[H_{\mathbf{k}}^1,H_{\mathbf{k}}^{-1}\right] \nonumber \\
&&+\frac{\left[H_{\mathbf{k}}^{-1},\left[H_{\mathbf{k}}^0,H_{\mathbf{k}}^1\right]\right]+\left[H_{\mathbf{k}}^1,\left[H_{\mathbf{k}}^0,H_{\mathbf{k}}^{-1}\right]\right]}{2
\Omega ^2} \nonumber \\
&&-\frac{\left[H_{\mathbf{k}}^1,\left[H_{\mathbf{k}}^{-2},H_{\mathbf{k}}^1\right]\right]+\left[H_{\mathbf{k}}^{-1},\left[H_{\mathbf{k}}^2,H_{\mathbf{k}}^{-1}\right]\right]}{3\Omega
^2}+\frac{\left[H_{\mathbf{k}}^{-1},\left[H_{\mathbf{k}}^{-1},H_{\mathbf{k}}^2\right]\right]+\left[H_{\mathbf{k}}^1,\left[H_{\mathbf{k}}^1,H_{\mathbf{k}}^{-2}\right]\right]}{6\Omega
^2},
\label{eq: ham_eff}
\end{eqnarray}
\end{widetext}
where \(H_{\mathbf{k}}^n\) is computed from Eq.(\ref{eq: ham_n}). We will discuss the form of \(H_{\mathbf{k}}^{\text{eff}}\) in different polarization of the laser field.

\subsection{Circularly polarized laser field}

In the circularly polarized light, one has the following Fourier components: 

\begin{eqnarray}
H_{\mathbf{k}}^0&=&3t p_x\sigma _y+\left(\frac{3t}{4}p_y{}^2+\frac{9t}{4}p_x{}^2\right)\sigma _x, \nonumber \\
H_{\mathbf{k}}^1&=&\left(\frac{3t A p_x}{4} -i\frac{3t A p_y}{4}\right)\sigma _x+\frac{3t A}{2}\sigma _y,\nonumber \\
H_{\mathbf{k}}^{-1}&=&\left(\frac{3t A p_x}{4}+i\frac{3t
A p_y}{4}\right)\sigma _x+\frac{3t A}{2}\sigma _y, \nonumber \\
H_{\mathbf{k}}^2&=&-\frac{3t A^2}{8}\sigma _x,H_{\mathbf{k}}^{-2}=-\frac{3t A^2}{8}\sigma _x,
\label{eq: ham_n_circ}
\end{eqnarray}
By inserting Eq.(\ref{eq: ham_n_circ}) into Eq.(\ref{eq: ham_eff}), 
\begin{widetext}
\begin{equation}
H_{\mathbf{k}}^{\text{eff}}=\left(\frac{3t}{4}p_y{}^2+\frac{9t}{4}p_x{}^2-\frac{27 t^3A^2}{8 \Omega ^2} \left(p_x{}^2+p_y{}^2+A^2\right)\right)\sigma
_x+3t p_x\left(1-\frac{9 t^2A^2}{8 \Omega ^2}\left(p_x{}^2-p_y{}^2+\frac{A^2}{2}\right)\right)\sigma _y+\frac{9(t A)^2}{4\Omega }p_y\sigma _z.
\label{eq: ham_eff_circ}
\end{equation}
\end{widetext}
The energies of Eq.(\ref{eq: ham_eff_circ}) contain a gap \(\Delta =\frac{27 t^3A^4}{4\Omega ^2}\) at \((0,0)\). Notice that we keep \(p_x{}^2\) term only for the convenience of momentum expansion. The dispersion is dominated by \({\cal O}\left(k_x\right)\). 

\subsection{Linearly polarized laser field}
In the linearly polarized field along $x$-direction, the effective Hamiltonian reads
\begin{equation}
H_{\mathbf{k}}^{\text{eff}}=H_{\mathbf{k}}^0=\left(\frac{3 t}{4}p_y{}^2 +\frac{9t}{4}p_x{}^2-\frac{3t}{8}A^2\right)\sigma _x+3t p_x\sigma _y.
\label{eq: ham_eff_linx}
\end{equation}
The spectrum of  Eq.(\ref{eq: ham_eff_linx}) includes two symmetric Dirac points along $y$-direction. The distance between the two band touching points is \(|\Delta \mathbf{k}|=\sqrt{2}A\).
In the linearly polarized field along $y$-direction, the effective Hamiltonian reads
\begin{equation}
H_{\mathbf{k}}^{\text{eff}}=H_{\mathbf{k}}^0=\left(\frac{3 t}{4}p_y{}^2 +\frac{9t}{4}p_x{}^2+\frac{3t}{8}A^2\right)\sigma _x+3t p_x\sigma _y,
\end{equation}
which contains a gap of size \(\Delta =\frac{3t A^2}{4}\).

\end{document}